\documentclass[a4paper,11pt]{article}

\usepackage{jcappub} 

\usepackage[T1]{fontenc}

\usepackage[english]{babel}
\usepackage{xparse}
\usepackage{comment}

\title{Multi-TeV dark matter density in the inner Milky Way halo: spectral and dynamical constraints}
\author[a, b]{Jaume Zuriaga-Puig,}
\author[a, b]{Viviana Gammaldi,}
\author[c]{Daniele Gaggero,}
\author[d]{Thomas Lacroix}
\author[a,b]{and M. A. Sánchez-Conde}

\affiliation[a]{Instituto de Física Teórica, IFT UAM-CSIC,\\ Calle Nicolás Cabrera 13-15, Campus de Cantoblanco, E-28049 Madrid, Spain.}
\affiliation[b]{Departamento de Física Teórica, Mod. 15, Universidad Autónoma de Madrid, \\ E-28049 Madrid, Spain.}
\affiliation[c]{INFN Sezione di Pisa, Polo Fibonacci, Largo B. Pontecorvo 3, 56127 Pisa, Italy.}
\affiliation[d]{5 B rue Francisque Jomard, 69600 Oullins, France.}
\emailAdd{jaume.zuriaga@csic.es}
\emailAdd{viviana.gammaldi@uam.es} 
\emailAdd{daniele.gaggero@pi.infn.it} 
\emailAdd{thomas.lacroix@lupm.in2p3.fr}
\emailAdd{miguel.sanchezconde@uam.es} 

\abstract{We develop a comprehensive study of the gamma-ray flux observed by the High Energy Stereoscopic System (H.E.S.S.) in 5 regions of the Galactic Center (GC). Motivated by previous works on a possible Dark Matter (DM) explanation for the TeV cut-off observed by H.E.S.S. in the innermost $\sim 15$ pc of the Galaxy, we aim to constrain the DM distribution up to a radius of $\sim 450$ pc from the GC. In this region, the benchmark approach (e.g. cosmological simulations and Galactic dynamics studies) fails to produce a strong prediction of the DM profile. Within our proof-of-concept analysis, we use DRAGON to model the diffuse background emission and determine upper limits on the density distribution of thermal multi-TeV Weakly Interactive Massive Particles (WIMPs), compatible with the observed gamma-ray flux. The results are in agreement with the hypothesis of an enhancement of the DM density in the GC with respect to the benchmark Navarro-Frenk-White (NFW) profile ($\gamma=1$) and allow us to exclude profiles with an inner slope cuspier than $\gamma \gtrsim 1.3$. We also investigate the possibility that such an enhancement could be related to the existence of a DM spike associated with the supermassive black hole Sgr A* at the GC. We find out that the existence of an adiabatic DM spike smoothed by the scattering off of WIMPs by the bulge stars may be consistent with the observed gamma-ray flux if the spike forms on an underlying generalized NFW profile with $\gamma \lesssim 0.8$, corresponding to a spike slope of $\gamma_{sp-star}= 1.5$ and spike radius of $R_\text{sp-stars} \sim 25$-$30$ pc. Instead, in the extreme case of the instantaneous growth of the black hole, the underlying profile could have up to $\gamma \sim 1.2$, a corresponding $\gamma_{sp-inst}=1.4$ and $R_\text{sp-inst}\sim 15$-$25$ pc. Finally, the results of our analysis of the total DM mass enclosed within the S2 orbit (updated with new GRAVITY data) are less stringent than the spectral analysis.  Our work aims to guide future studies of the GC region, with both current and next generation of telescopes. In particular, the next Cherenkov Telescope Array will be able to scan the GC region with improved flux sensitivity and angular resolution.
}

\keywords{dark matter distribution, gamma rays, Galactic Center, WIMPs}

\begin{document}

\maketitle

\flushbottom

\newpage

\section{Introduction}
\label{introduction}

Many observations, such as galaxy rotation curves \cite{Benito20}, gravitational lensing \cite{2006ApJ...648L.109C} and the cosmic microwave background \cite{Aghanim:2018eyx} have led to the conclusion that non-baryonic Dark Matter (DM) constitutes about the $84\%$ of the total mass content of the universe, yet its nature is still unknown. Among other candidates \cite{2012LRR....15...10F}, cold DM is able to explain most of the astrophysical and cosmological evidence, with one natural candidate being the Weakly Interactive Massive Particles (WIMPs). Many WIMP candidates are expected to have been produced thermally in the early Universe, similarly to the particles of the Standard Model (SM), and usually constitute cold DM. Obtaining the correct abundance of DM today via thermal production requires a self-annihilation cross-section between $\langle \sigma v\rangle \simeq 5.2 \times 10^{-26}\mathrm {cm} ^{3}\;\mathrm {s} ^{-1}$ at $\approx 0.3$ GeV and $\langle \sigma v\rangle \simeq 2.2 \times 10^{-26}\mathrm {cm} ^{3}\;\mathrm {s} ^{-1}$ above $\approx 10$ GeV \cite{Steigman:2012nb}. Many efforts have been addressed to detect WIMPs, being focused mainly on searches at colliders \cite{2022PhRvD.105i2007T}, direct detection experiments \cite{2022arXiv220308084A} and indirect searches \cite{Gaskins:2016cha, MiniReview}. SM particles are accelerated in colliders, where they could produce, among other particles, DM particles which can be detected as a ``missing energy'' in the original process. For the direct detection experiments, the transferred energy from the DM particle to SM particles via elastic collisions is expected to be detectable. In this work, we focus on indirect searches, which rely on the detection of secondary fluxes of astroparticles produced in the annihilation/decay of DM in astrophysical targets. Those fluxes can be observed in experiments and observatories such as High Energy Spectroscopic System (H.E.S.S.), Major Atmospheric Gamma-Ray Imaging Cherenkov (MAGIC), High-Altitude Water Cherenkov Observatory (HAWC), Fermi, etc., allowing to set constraints on a broad range of WIMP masses and the associated parameter space (annihilation cross-section or decay time).

Astrophysical targets of interest for indirect detection of DM are traditionally dwarf galaxies \cite{Charles:2016pgz, 2018PhRvD..98h3008G, 2021PhRvD.104h3026G, 2022PDU....3500912A, 2020PhRvD.102f2001A}, the Galactic Center (GC) \cite{Di_Mauro_2021, Acharyya_2021, 2012PhRvD..86j3506C, Cembranos_2013, PhysRevLett.129.111101, 2022PhRvL.129w1101Z}, or  Galaxy Clusters \cite{2023PhRvD.107h3030D, 2012ApJ...750..123A}. The most important features of an astrophysical target are its DM component, its distance, and all the possible uncertainties relating to the modeling of the object, like the background flux. Dwarf galaxies are desirable targets since they are DM-dominated systems, with a DM mass of $10^7$-$10^{10}$ $\text{M}_{\odot}$ and a negligible astrophysical background, that are close to the earth ($\sim 0.5$ Mpc). Focusing on specific targets, dwarf spheroidal galaxies have the counterpart of its dynamics being pressure-dominated, making it difficult to estimate its DM density profile. Dwarf irregular galaxies are, on the other hand, more distant objects but dominated by rotation, alleviating the problem (e.g., see \cite{Charles:2016pgz, 2018PhRvD..98h3008G}). Galaxy Clusters are DM-dominated as well, although they are distant objects with high astrophysical backgrounds, so propagation effects need to be modeled. Finally, the GC is the closest source to the Earth with the highest expected DM annihilation flux. However, the GC has a rich astrophysical background, both from sources and diffuse galactic emission, whose modeling can be a challenging task.

Besides the background modeling, one of the highest sources of uncertainty in the indirect detection of DM is modeling the DM density distribution profile in the astrophysical targets. The DM distribution has been a subject of debate for years. On one hand, DM-only numerical simulations favor steeper profiles, generally well described by the benchmark Navarro-Frenk-White profile (NFW, \cite{NFW}). Hydrodynamical simulations, which also include the effect of baryons \cite{1990ApJ...356..359H}, seem to favor a generalized NFW profile with a different slope contracting the inner halo with higher densities \cite{2020MNRAS.494.4291C, 2022MNRAS.513...55M, 2022MNRAS.511.3910F} or with the existence of the bar in the Galaxy that could also affect the shape of the halo \cite{2022MNRAS.511.3910F}. On the other hand, observations of the rotation curves of dwarf galaxies prefer cored profiles \cite{2018PhRvD..98h3008G, 2021PhRvD.104h3026G}. Regarding our Galaxy, the outer shape of the profile is mostly constrained by the observed rotation curves \cite{Benito20}, whose results put constraints on the parameters that define it. Nonetheless, difficulties in obtaining dynamical measurements of the baryonic matter (due to the fact that we are inside the Milky Way) translate into difficulties when determining the distribution of the DM  at a distance $\lesssim 2.5 $-$ 3$ kpc from the GC \cite{2015A&A...578A..14C, 2017PDU....15...90I}. Despite this issue, upper limits on the DM mass distribution in the very central regions of the GC ($< 10^{-2}$ pc) can be obtained by observing the orbits of the S stars surrounding the Super Massive Black Hole (SMBH) Sgr A* \cite{GRAVITY_2021, GRAVITY_2020, Do2019, Lacroix_2018}. When considering parsec scales, the uncertainty on the DM distribution is even worse. In fact, a possible enhancement of the DM density (hereafter, DM spike) could be associated with the growth of the central SMBH Sgr A*. This spike may have different characteristics depending on the evolution history of the SMBH, e.g. the adiabatic growth of the SMBH \cite{PhysRevLett.83.1719, Sadeghian2013}, the interaction with the stars surrounding the central spike \cite{PhysRevD.78.083506, doi:10.1142/S0217732305017391}, the extreme case of the instantaneous growth of the SMBH \cite{Ullio2001} or the effects of a rotating Kerr BH \cite{Ferrer2017}. The biggest question regarding DM spikes is their stability and survival in galaxies such as the Milky Way.

In this work, we aim to set new constraints on the DM distribution in our Galaxy. We develop a comprehensive study of compatibility between the analysis of the gamma-ray spectra observed by H.E.S.S. in 5 different concentric regions in the GC \cite{HESSII, Collaboration2018, HESS:2017tce, HESSRidge, HESSHalo, PhysRevLett.106.161301, PhysRevLett.129.111101}, the dynamical constraints both in the outer \cite{Benito20, McMillan2017} and inner region \cite{Lacroix_2018, Shen2023} of the Galaxy, and the possibility of having a detected multi-TeV DM signature in the inner GC \cite{HESSI, 2012PhRvD..86j3506C, Cembranos_2013}. Despite the benchmark approach in DM indirect searches, which aims to set constraints on the DM annihilation cross-section by selecting a couple of possible (typically, one core and one cusp) DM density distribution profiles (see e.g. \cite{Montanari_2022buj, 2023JCAP...08..063B}), our main focus is to set constraints on the DM density profile within in a radius of 450 pc from the GC, by assuming a thermal DM candidate which could explain the gamma-ray spectra detected by H.E.S.S. in the inner 15 pc of the Galaxy \cite{2012PhRvD..86j3506C, Cembranos_2013}. Our hypothesis is based on the fact that when considering a multi-TeV DM candidate with a thermal cross-section $\langle\sigma v\rangle = 3 \times 10^{-26}$ cm$^3$s$^{-1}$, a boost factor is needed in the DM density profile (assumed to be a benchmark NFW) to explain the gamma-ray flux observed by H.E.S.S. in the source J1745-290 at the Very Inner Region (VIR) of the Galaxy \cite{2012PhRvD..86j3506C, Cembranos_2013, Gammaldi_2016}. Furthermore, the angular dimension of such local enhancement of the DM density could favor the disentangling between the existence of a cusp profile or a DM spike. This can be summarized as, from a morphological point of view, that the HESS J1745-290 data would be compatible with the scenario of a continuous DM annihilation signal.

The manuscript is organized as follows. In Section \ref{GC} we introduce the gamma-ray spectra observed by H.E.S.S. in 5 concentric regions at the GC and its possible interpretation as DM signal. In Section \ref{Spectral_analysis} we discuss the new background modeling, the spectral analysis and the results obtained for the spectral constraints on the DM distribution. We compare these results with DM density distribution profiles obtained by both dynamical studies and simulations in Section \ref{J_factors_morphology}. We verify the compatibility of our results with the dynamical constraints obtained at sub-parsec scale by the analysis of the S2 star orbit in Section \ref{dynamical_constraints}. Finally, the conclusions can be found in Section \ref{conclusions}.

\begin{figure}[t!] 
  \centering 
  \includegraphics[width=0.49\textwidth]{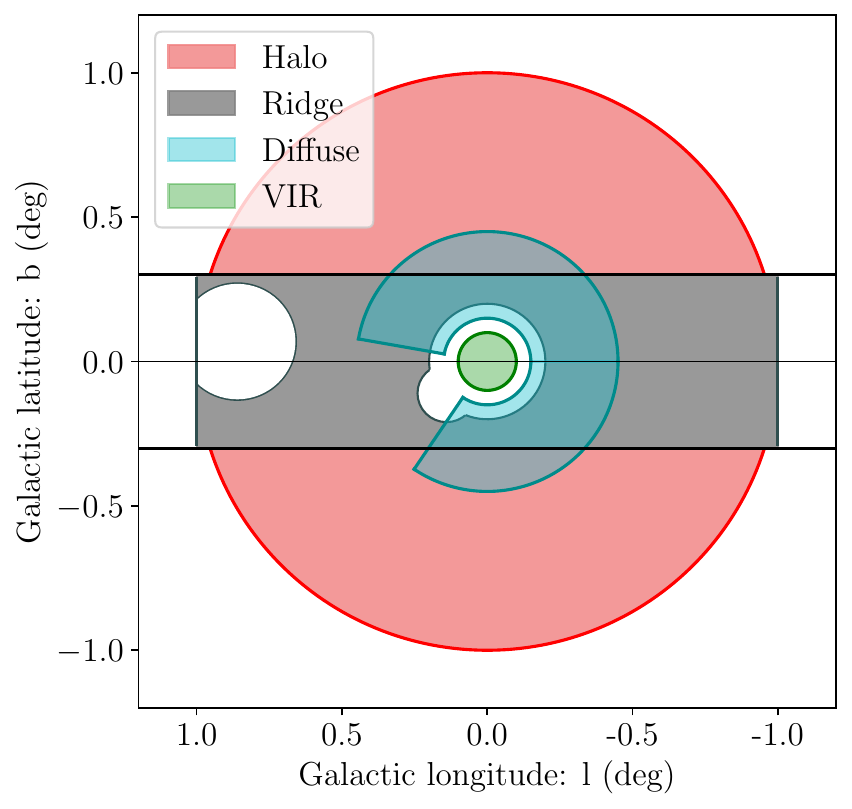} 
  \includegraphics[width=0.49\textwidth]{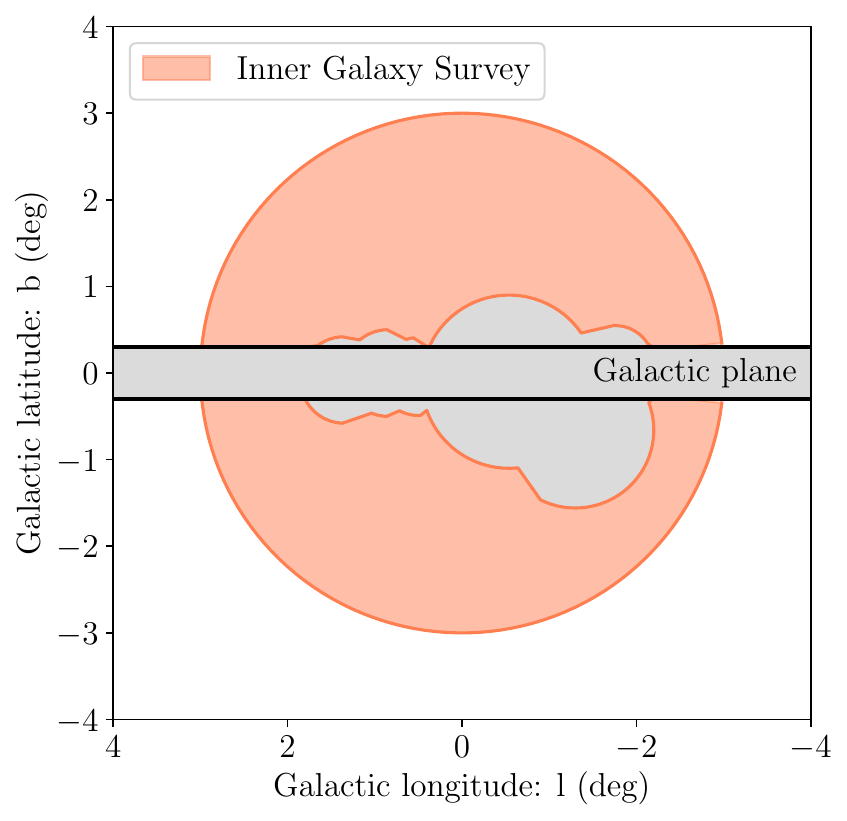} 
  \caption{\footnotesize{5 Regions of Interest. Left panel: VIR, $\theta < 0.1^\circ$, r $\lesssim$ 15 pc (in green in the figure); Ridge, $|b| < 0.3^\circ$ (43 pc) and $|l| < 1.0^\circ$ (145 pc) (in gray); Diffuse region, $0.15^\circ < \theta < 0.45^\circ $, 22 pc $\lesssim$ r $\lesssim$ 65 pc (in blue); Halo, $0.3^\circ < \theta < 1.0^\circ$, 43 pc $\lesssim$ r $\lesssim$ 145 pc (in red), excluding the latitudes $|b| < 0.3^\circ$ (the Galactic plane). Right panel: IGS, $0.5^\circ<\theta<3.0^\circ$, 72 pc $\lesssim$ r $\lesssim$ 434 pc (in orange), excluding the Galactic plane and other complex regions (light grey).}}
  \label{fig:regions_GC} 
\end{figure}

\section{The Galactic Center region}
\label{GC}

As anticipated in the introduction, the GC region is still a hot topic for the indirect detection of DM. Due to the complexity of the region, distinguishing the DM signal and the astrophysical background is a very hard issue \cite{Gaggero:2017jts}. So far, there are two main research directions dealing with a prospective DM signal at the GC: the first one is the GeV excess detected by Fermi-LAT \cite{Di_Mauro_2021, 2016JCAP...05..049B, Calore:2014hna, PhysRevD.91.063003}, the second one is the TeV cut-off by H.E.S.S. \cite{2012PhRvD..86h3516B, MiniReview, Cembranos_2013, 2012PhRvD..86j3506C, 2014PhRvD..90d3004C, Cembranos:2014wza}.  While the first hypothesis has been extensively studied in the literature, we focus our study on the multi-TeV DM candidate for the gamma-ray cut-off detected by H.E.S.S. In this section, we introduce the collection of H.E.S.S. gamma-ray data in the GC region, the indirect detection of WIMP particles, the multi-TeV WIMP candidate and the open issue about the DM density distribution profile.

\subsection{H.E.S.S. spectral data}
\label{ROI}

Thanks to its position in the southern hemisphere, the H.E.S.S. telescope is currently the only Imaging Air Cherenkov Telescope capable of observing the GC at TeV energy with good angular and energy resolution.  For this reason, we collect the H.E.S.S. published data of 5 concentric regions of the GC, with the aim to perform a spatial and spectral study of the gamma-ray flux in the region. The collection of 5 regions analyzed in this work is shown in Figure \ref{fig:regions_GC}: the Very Inner Region (VIR) \cite{HESSI, HESSII}, the Ridge \cite{HESS:2017tce, HESSRidge}, the Diffuse emission region \cite{HESSII}, the Halo \cite{HESSHalo, PhysRevLett.106.161301} and the Inner Galaxy Survey (IGS) \cite{PhysRevLett.129.111101}. General spectral and spatial information of each region is given by the H.E.S.S. collaboration, by fitting the gamma-ray flux with the following function:

\begin{equation}
  \frac{d \Phi_{\text{total}}}{d E}=\Phi_\text{point} (\frac{E}{1 \text{TeV}})^{-\Gamma} e^{-(\frac{E}{E_{\text{cut}}})}.
  \label{eq:fit_HESS} 
\end{equation}

These values are summarized in the following lines:

\paragraph{H.E.S.S. J1745-290 VIR.}The very inner part of the GC region $\theta<0.1^\circ$ (Figure  \ref{fig:regions_GC}, green region),  with $\Delta\Omega=9.57\times10^{-5}$ sr. A power-law with $\Phi_\text{VIR}=2.55\pm0.41\times10^{-12}\text{ TeV}^{-1}\text{cm}^{-2}\text{s}^{-1}$, $\Gamma_\text{point}=2.14\pm0.12$ and a preference for an exponential cut-off at $E_\text{cut}=10.7\pm4.1$ TeV is fitted \cite{HESSII}. 

\paragraph{H.E.S.S. Ridge.}The region $|b|< 0.3^\circ$ and $|l|< 1.0^\circ$ is studied (in dark grey, Figure \ref{fig:regions_GC}), with $\Delta \Omega = 3.26 \times 10^{-4}$ sr. Such a region shows a gamma-ray flux that at low energies is comparable with the cosmic-ray (CR) diffusion model, with $\Phi_\text{Ridge}=1.2\pm0.24\times10^{-11}\text{ TeV}^{-1}\text{cm}^{-2}\text{s}^{-1}$ and $\Gamma_\text{Ridge}=2.28\pm0.23$, and no cut-off detected.  At tens of TeV, it shows a deviation from the diffusion model. Note that the following masks are applied to the region: $0.2^\circ$ around Sgr A* and G0.9+0.1, and $0.1^\circ$ around HESS J1746-285. See \cite{HESS:2017tce} for more details.

\paragraph{H.E.S.S. Diffuse.}In an intermediate region (in blue, $0.15^\circ<\theta<0.45^\circ$, with $\Delta\Omega=1.41\times10^{-4}$ sr) it was detected a diffuse emission whose best fit is a power-law signal with $\Phi_\text{Diff}=1.92\pm0.36\times10^{-12}\text{ TeV}^{-1}\text{cm}^{-2}\text{s}^{-1}$ and index $\Gamma_\text{diff}=2.32\pm0.16$, and no evidence of any cut-off. A region of $66^\circ$ is excluded from the disc, which is bounded by the opening angles of $10^\circ$ and $-56^\circ$ taken from the positive galactic longitude axis \cite{HESSII}.
  
\paragraph{H.E.S.S. Halo.} In the outer region (in red, $0.3^\circ<\theta<1.0^\circ$, with $\Delta\Omega=5.97\times10^{-4}$ sr) no signal is detected when subtracting the background emission from a symmetric OFF region \cite{HESSHalo, PhysRevLett.106.161301}. The galactic latitudes $|b|< 0.3^\circ$ (almost coinciding with the Ridge region) are excluded from the analysis.

\paragraph{H.E.S.S. Inner Galaxy Survey (IGS).} Figure \ref{fig:regions_GC} (right panel, in orange), the region is defined as $0.5^\circ<\theta<3.0^\circ$, with $\Delta\Omega=6.38\times10^{-3}$ sr. Such as in the Halo, no signal is detected when subtracting the background emission from a symmetric OFF region. The galactic latitudes $|b|< 0.3^\circ$ are excluded from the analysis, as well as other regions (in grey in the figure) \cite{PhysRevLett.129.111101}. For the Halo and IGS regions, in the ON-OFF analyses no excess is found. Instead, a Test Statistics analysis is performed to compute the corresponding upper limits.

\vspace{5mm} 

In other words, a cut-off in the gamma-ray flux (compatible with a TeV DM annihilation signal, \cite{HESSII}) has been detected by the H.E.S.S. collaboration only in the VIR. From a morphological point of view, we hypothesize that it would be compatible with the scenario of a continuous DM annihilation signal. If a DM cusp exists at the GC, we may expect that in the inner region the annihilation component of the gamma-ray flux becomes more important over the background than in the external regions, where the DM detection could be locally suppressed by the astrophysical background. In the following sections, we study in-depth the gamma-ray flux in each region in order to model the DM density distribution in the Galaxy of the possible multi-TeV WIMP candidate, which well fits the gamma-ray cut-off in the VIR.

\subsection{Gamma-ray flux from WIMP annihilation} 
\label{Indirect_Detection_Gamma_Rays}

The gamma-ray flux produced by WIMP annihilation has the following form:

\begin{equation}
  \frac{d \Phi_{\text{DM}}}{d E}=\sum_{i}^{\text {channels }} \frac{\langle\sigma v\rangle_{i}}{2}  \frac{d N_{i}}{d E}  \frac{\Delta \Omega\langle J\rangle_{\Delta \Omega}}{4 \pi m_{\mathrm{DM}}^{2}}
  \label{eq:dflux/de} 
\end{equation}

Where $\langle\sigma v\rangle_{i}$ is the thermally averaged annihilation cross-section for the WIMP particle, $m_{\text{DM}}$ is the mass of the DM particle, and $\Delta \Omega$ is the solid angle of the observed region. In a model-independent approach, the WIMP particle is considered to annihilate in only one i-th channel of the Standard Model (SM). Also, by assuming a specific WIMP candidate with different branching ratios, the annihilation could happen in a combination of several SM channels. In particular, brane-world DM is an interesting possibility to naturally produce thermal multi-TeV WIMP candidates \cite{PhysRevLett.90.241301, https://doi.org/10.48550/arxiv.2205.07055, 2020JCAP...10..041A, 2020PDU....2700448C}. Both those possibilities have been studied in \cite{Cembranos_2013, 2012PhRvD..86j3506C, 2020JCAP...10..041A, 2014PhRvD..90d3004C}. Motivated by those previous studies, we focus on the ZZ channel, which gives one of the best fits to the H.E.S.S. data in the inner 15 pc of the GC. The factor $\frac{dN_i}{dE}$ is the differential flux of secondary particles, in this case gamma rays, produced by the possible hadronizations, decays, annihilations and other interactions that can create them starting from the primary products of DM annihilation, indeed the SM annihilation channel. They are computed using PPPC4DMID \cite{Cirelli_2011, Ciafaloni2010}. It can be shown that including interactions on the secondary particles created by DM annihilation such as electrons/positrons (Inverse Compton, Bremsstrahlung, and synchrotron radiation) the total DM spectra can change, but since we are in an energy range of $\sim 10^2$-$10^5$ GeV this flux is a few orders of magnitude below the primary gamma-ray flux \cite{Djuvsland2022, Buch2015}, so it is a good approximation to neglect them.  Note that, for kinematic reasons, the energy of the gamma rays generated by DM annihilation cannot exceed its own mass.

Finally, the J-factor $\langle J\rangle_{\Delta \Omega}$ (also called astrophysical factor in the literature) is where the information about DM distribution profile is contained \cite{Evans_2004}:

\begin{equation}
  \langle J\rangle_{\Delta \Omega}=\frac{1}{\Delta \Omega} \int_{\Delta \Omega} \mathrm{d} \Omega \int_{l(\hat{\theta})_{\min }}^{l(\hat{\theta})_{\max }} \rho^{2}_{\text{DM}}[r(l)] d l(\hat{\theta})
  \label{eq:astrophysicalfactor} 
\end{equation}

The J-factor is the integral of the DM density profile squared $\rho_{\text{DM}}(r)$, along the line of sight $l$, and averaged over the solid angle $\Delta \Omega$. In these coordinates, $l(\hat{\theta})_{\min}$ and $l(\hat{\theta})_{\max }$ are the edges of the regions observed in the direction given by $\hat{\theta}$ (if the region corresponds to the integration starting at the position of the Sun, $l(\hat{\theta})_{\min} = 0$). $r$ is the radial coordinate from the center of the GC (taken in Sgr A*), and can be related to the line of sight $l$ with the expression $r^2 = l^2 + R^2_{\odot} - 2R_{\odot} l \cos(\hat{\theta})$, where $R_{\odot} = 8.277$ kpc is the distance from the Sun to the Sgr A* (\cite{GRAVITY_2021}, hereafter GRAVITY2021). Assuming a virial radius $R_{\text{vir}}$ for the DM halo, we have that  $l(\hat{\theta})_{\max } = R_{\odot}\cos(\hat{\theta}) + \sqrt{R_{\text{vir}}^2-R^2_{\odot}\sin(\hat{\theta})^2}$.

\subsection{The multi-TeV WIMP candidate}
\label{Multi_TeV}

According to \cite{2012PhRvD..86j3506C, Cembranos_2013}, we assume here that the total gamma-ray flux detected by H.E.S.S. in the GC region is a combination of the gamma-ray flux produced by self-annihilating DM particles $\frac{d \Phi_{\mathrm{DM}}}{d E}$ (Equation \ref{eq:dflux/de}) and a background component $\frac{d \Phi_{\text{Bg}}}{d E}$:

\begin{equation}
  \frac{d \Phi_{\mathrm{total}}}{d E}=\frac{d \Phi_{\mathrm{DM}}}{d E} + \frac{d \Phi_{\text{Bg}}}{d E}
  \label{eq:dflux/de-total} 
\end{equation}

This hypothesis is well motivated by the difficulty with modeling the astrophysical background emission in the region, due to the abundance of known and unknown astrophysical sources because of the particularly bright diffuse emission caused by the cosmic-ray interactions with the interstellar medium. We allow the normalization of the background to vary for matching the data and allow significant fluctuations of this normalization in the very small regions (such as the VIR) with respect to the Ridge region, motivated by a combination of uncertainties on modeling the column density of the molecular gas and on the properties of the CR sea at small scales.

By assuming a benchmark NFW profile, a boost factor of $\sim 10^3$ in the J-factor was needed in order to explain the observed signal in the inner 15 pc of the GC, with a DM mass $m_{\text{DM}} \simeq 50$ TeV \cite{2012PhRvD..86j3506C, Cembranos_2013}. On the one hand, such a boost factor may be compatible with the characteristics expected by an enhancement of the DM distribution in the region due to the presence of the SMBH Sgr A* \cite{Gammaldi_2016}. Depending on the DM distribution profile in the Galaxy (cusp or core), both the slope and the radius of the DM spike change, by giving a footprint on the inner DM distribution and, indeed, on the morphology of the expected gamma-ray emission due to the annihilation of WIMPs \cite{Gammaldi_2016}. On the other hand, the spectral cut-off strictly depends on the DM mass and the background component. To derive these results, the authors used a background given by a simple power law ($\frac{d \Phi_{\text{Bg}}}{d E} \propto E^{-\Gamma}$) \cite{2012PhRvD..86j3506C, Cembranos_2013}. Our work is based on these results but with the novelty of using the state-of-the-art background model computed with the public numerical codes DRAGON and HERMES  \cite{Evoli_2017, 2017JCAP...10..019C} and extending the analysis up to 5 concentric regions in the GC.

\subsection{The DM density profile}
\label{morphology}

\begin{figure}[t!] 
  \centering 
  \includegraphics[width=9cm]{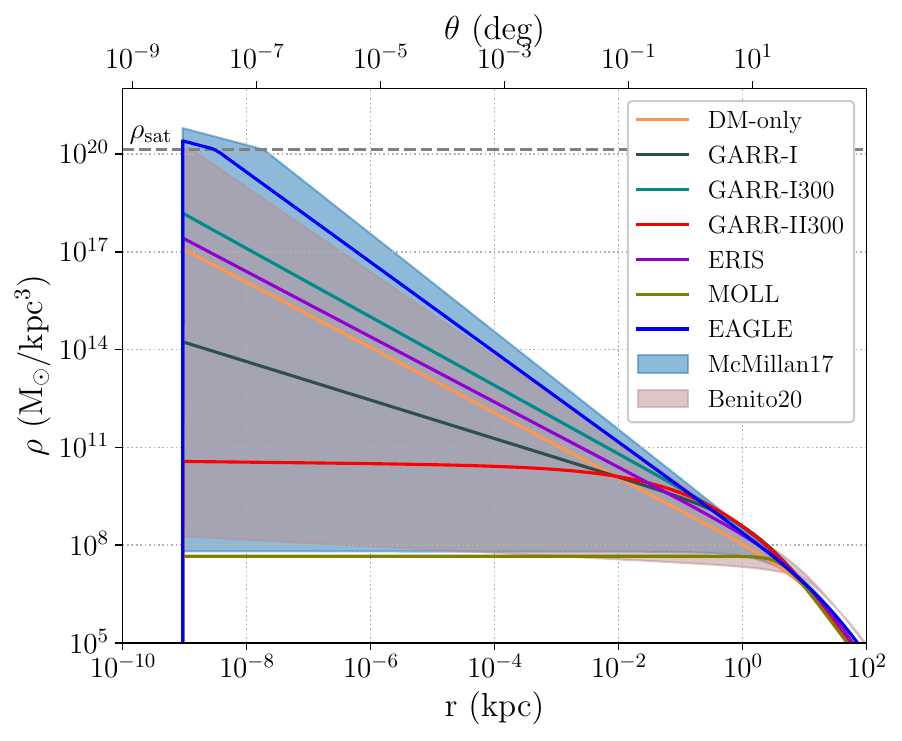} 
  \caption{\footnotesize{DM density distribution profiles considered in this work. Profiles represented with a solid line correspond to cosmological simulations: DM-only (orange line), GARR-I (gray), GARR-I300 (light blue), GARR-II300 (red), ERIS (purple), MOLL (green) and EAGLE (blue). The blue (McMillan17) and brown (Benito20) regions cover the range of possible profiles which are compatible with the study of Galactic dynamics (see main text for details). To ensure an easy reading, we kept this color code along the manuscript. All the profiles are extrapolated up to the inner radius of the Galaxy, defined by the Schwarzschild radius (see Section \ref{J_factors_Spike_Stars} for further details).  The grey horizontal dashed line represents the central plateau due to DM annihilation $\rho_{\text{sat}} = m_{\text{DM}} / (t_{\text{BH}} \langle\sigma v\rangle)$ = $1.4\times 10^{20} \text{ M}_{\odot} \text{kpc}^{-3}$ for a mass of $m_{\text{DM}} = 36$ TeV (see text for further details). The parameters of the profiles can be found in Table \ref{tab:density_parameter_table}.}}
  \label{fig:density_models} 
\end{figure}

\begin{center}
\begin{table}[t!]
\begin{center}
\resizebox{\textwidth}{!}
{
\begin{tabular}[b]{|c|c|c|c|c|c|c|}

\hline
\hline
Profile & $\gamma$ & $\alpha$ & $\beta$ & $\rho_s\left(\text{M}_\odot \text{kpc}^{-3}\right)$ & $r_{s}\left(\text{kpc}\right)$ & $\rho_\odot(\text{GeV}\text{cm}^{-3})$ \\
\hline
\hline

DM-only & $1$ & $1$ & $3$ & $5.38\times10^6$ & $21.5$ & $0.28$ \\
\hline
GARR-I & $0.59$ & $1$ & $2.70$ & $4.97\times10^8$ & $2.3$ & $0.35$ \\
\hline
GARR-I300 & $1.05$ & $1$ & $2.79$ & $1.01\times10^8$ & $4.6$ & $0.35$ \\
\hline
GARR-II300 & $0.02$ & $0.42$ & $3.39$ & $2.40\times10^{10}$ & $2.5$ & $0.35$ \\
\hline
ERIS & $1$ & $1$ & $3$ & $2.25\times 10^7 $ & $10.9$ & $0.36$ \\
\hline
MOLL & $8 \times 10^{-9}$ & $2.89$ & $2.54$ & $4.57\times 10^{7}$ & $4.4$ & $0.31$ \\
\hline
EAGLE & $1.38$& $1$ & $3$ & $2.18\times10^{6}$ & $31.2$ & $0.35$ \\
\hline
McMillan17 & $0$-$1.5$ & $1$ & $3$ & $1.2\times10^{8}$-$5.3\times10^{5}$ & $6.8$-$59.9$ & $0.33$-$0.43$ \\
\hline
Benito20 & $0.1$-$1.3$ & $1$ & $3$ & $1.8\times10^{8}$-$2.5\times10^{6}$ & $7.0$-$40.0$ & $0.41$-$0.71$ \\
\hline

\end{tabular}
}
\end{center}
\caption{\footnotesize{Parameters of different DM density profiles as in Equation \ref{eq:NFW_general} for the DM-only simulation, the hydrodynamical simulations GARR, GARR-I, GARR-I300, GARR-II300, ERIS, MOLL and EAGLE (see \cite{Gammaldi_2016} and the references within) and the observational models McMillan17 \cite{McMillan2017} and Benito20 \cite{Benito20}. Also, it is shown the local DM density $\rho_\odot$ of each simulation, with $R_\odot = 8.277$ kpc (GRAVITY2021 \cite{GRAVITY_2021}). 
}}
\label{tab:density_parameter_table}
\end{table}
\end{center}

As introduced in Section \ref{Indirect_Detection_Gamma_Rays}, high uncertainty exists in determining the DM density distribution profiles in the Galaxy. Of course, this fact has a high impact on studying prospective DM annihilation signals. In this work, we consider a wide range of DM density profile models, from cuspy to cored profiles and from both cosmological simulations and dynamical observations. Here, we adopt the Navarro-Frenk-White (NFW) generalized formalism \cite{Zhao:1995cp}: 

\begin{equation}
  \rho_{\text{halo}}(r) = \frac{ \rho _s}{(\frac{r}{r_{s}})^{\gamma} (1+(\frac{r}{r_{s}})^{\alpha})^{\frac{\beta - \gamma}{\alpha}} }, 
  \label{eq:NFW_general} 
\end{equation}

where $\rho_s$ is a normalizing factor and $r_{s}$ is the scale radius of the profile. This expression allows us to recover the benchmark NFW profile resulting from DM-only N-body simulations, assuming ($\alpha$, $\beta$, $\gamma$) = (1,3,1) \cite{NFW, Evans_2004}, or more complicated distribution obtained by hydrodynamical simulations \cite{1990ApJ...356..359H} (here, the GARR-I, GARR-I300 and GARR-II300 \cite{2016ApJ...824...94R}, ERIS \cite{2011ApJ...742...76G}, MOLL (Halo B in \cite{2015MNRAS.447.1353M}), EAGLE (Halo 1 in \cite{2016MNRAS.455.4442S}); for a discussion of the parameters, see \cite{Gammaldi_2016}). Under this formalism, we can also describe two models obtained from observations: 1) the set of NFW-like models for values of $\gamma$ between 0 and 1.5 \cite{McMillan2017} (hereafter, McMillan17); 2) the 2$\sigma$ constraints on the DM profile in the Milky Way obtained by the most recent analysis of the galactic rotation curve \cite{Benito20} (hereafter, Benito20\footnote{We refer to the 2$\sigma$ allowed region in the frequentist approach, which we use to stay conservative.}). For both the observational models, we have $\alpha=1$ and $\beta=3$. The parameters of each DM profile are given in Table \ref{tab:density_parameter_table} and the DM profiles themselves are shown in Figure \ref{fig:density_models}. From this figure, it is easily visible that - due to the high uncertainty in the data set - the range of parameters obtained for the DM profiles via the study of the galactic kinematic cover all the different profiles obtained via different cosmological simulations. For a better comparison of the density distribution profile parameters, see Section \ref{Appendix_params} in the Appendix. \\

Let us remark that neither simulations nor the study of the Milky Way rotation curve allows us to set constraints on the DM profile at parsec scales. In this work, we aim to set interesting upper limits on the DM density distribution in such an unconstrained region via the comprehensive analysis of the gamma-ray spectra detected by H.E.S.S. in the previously introduced 5 regions of the inner Galaxy (within 450 pc), by following the hypothesis of the existence of a DM component. Interestingly, this approach allows us to also set constraints not only on benchmark outer DM profiles (i.e. a core or cusp profile) but also on more complex models, which include possible modifications in the inner part of the outer benchmark DM profile, such as: 1) the existence of a DM-only adiabatic spike \cite{PhysRevLett.83.1719, Sadeghian2013}, 2) the effect of the interactions between stars and an adiabatic spike \cite{PhysRevD.78.083506, doi:10.1142/S0217732305017391}, 3) the extreme case of an instant spike \cite{Ullio2001}, and 4) the perturbation due to a rotating Kerr SMBH \cite{Ferrer2017}. Furthermore, the matter distribution in the Galaxy below $\lesssim 10^{-2}$ pc can be constrained by dynamical constraints of the S stars \cite{Lacroix_2018}. This fact will represent an additional independent cross-check to the prospective DM distribution modeled by the analysis of the gamma-ray spectra, as explained in Section \ref{dynamical_constraints}.\\

\section{Spectral analysis}
\label{Spectral_analysis}

Our analysis of the spectral data is based on a $\chi ^2$ statistical approach: 

\begin{equation}
  \chi^2 = \sum_{i=1}^{N} \frac{(E_i^2 y^{\mathrm{HESS}}_i - E_i^2 \frac{d \Phi_{\mathrm{total}}}{d E}(E_i))^2}{(E_i^2 y^{\mathrm{error}}_i)^2} 
  \label{eq:chi_2_formula} 
\end{equation}

where $E_i^2 y^{\mathrm{HESS}}_i$ and $E_i^2 y^{\mathrm{error}}_i$ are, respectively, the differential gamma-ray flux observed by H.E.S.S. and its uncertainty within the $E_i$ energy bin; $E_i^2 \frac{d \Phi_{\mathrm{total}}}{d E}(E_i)$ is the differential flux computed following Equation \ref{eq:dflux/de-total} and $N$ is the total amount of data points observed in the region. Therefore, the $ddof$ of this analysis would be the number of data points $N$ minus the number of the free parameters fitted.

\subsection{Background model}
\label{Background}

Beyond the uncertainty in modeling the DM density distribution profile, the second main issue in studying the GC region is the background flux modeling. Due to the presence of a very bright diffuse emission and several point sources, plus possibly a population of unresolved sources, separating the prospective DM annihilation signal from the astrophysical background emission is a hard task. 

The diffuse background is due to the presence of a diffuse charged cosmic rays (CR) ``sea'' in the Galaxy, effectively confined by its turbulent magnetic field \cite{Blasi:2013rva,Gabici:2019jvz, 2021NatCo..12.6169H}. The presence of this emission over a wide range of energies constitutes an unavoidable challenge whenever one tries to pinpoint any emission of exotic origin \cite{Gaggero:2018zbd}. Charged CRs interact with several components of the interstellar medium (interstellar gas, low-energy photons, regular and turbulent magnetic field) and emit a bright, diffuse radiation from the radio domain all the way up to the multi-TeV gamma-ray band as a consequence of a variety of non-thermal processes (mainly synchrotron, bremsstrahlung, neutral pion decay and Inverse Compton scattering). We model such background emission by means of the \textit{Diffusion Reacceleration and Advection of Galactic cosmic rays: an Open New code} (DRAGON) \cite{Evoli_2008,Evoli_2017,Evoli:2017vim, 2017JCAP...10..019C}, developed within the HERMES publicly available computational framework for the line of sight integration over galactic radiative processes, which creates sky maps in the HEALPix-compatible format \cite{Dundovic:2021ryb}. 
The diffuse model we adopt here is described in detail in a recent paper \cite{Luque:2022buq} and is tuned on different sets of local charged cosmic-ray, multi-wavelength and gamma-ray data along the Galactic plane. It features a harder diffusion coefficient (hence, a harder propagated proton spectrum) compared to the locally measured one, as hinted by Fermi-LAT data and first introduced in \cite{Gaggero:2014xla}. This feature was shown to be consistent with the bulk of the emission observed by the H.E.S.S. collaboration in the Galactic Ridge region \cite{Gaggero:2017jts}. Indeed, given this choice for the background, in Equation \ref{eq:dflux/de-total} we use 

\begin{equation}
  \frac{d \Phi_{\text{Bg}}}{d E} = B^2  \frac{d \Phi_{\text{DRAGON}}}{d E},
  \label{eq:DRAGON} 
\end{equation}

where $\frac{d \Phi_{\text{DRAGON}}}{d E}$ is obtained for each of the 5 regions by applying to the DRAGON sky map the same mask as in the H.E.S.S. data, and $B$ is an $\mathcal{O}(1)$ normalizing factor that is left as a free parameter. The reason why the overall normalization is allowed to vary is the existence of an intrinsic, unavoidable uncertainty in the model itself, which ultimately stems from our poor knowledge of the conversion factor between the CO emissivity and the molecular gas column density (see e.g. the recent review \cite{Tibaldo:2021viq}). The diffuse emission scales linearly with this poorly constrained parameter which we leave free to vary together with the exotic component originating from DM annihilation. 

\subsection{Methodology}
\label{Results_fit}

We assume a thermal WIMP particle with the benchmark annihilation cross-section $\langle\sigma v\rangle \simeq 2.2\times 10 ^{-26} \text{cm}^3 \text{s}^{-1}$ necessary to explain the observed DM relic abundance in the universe \cite{Steigman:2012nb}. Indeed we fit the following equation to the data of each region:

\begin{equation}
   \frac{d \Phi}{d E} = B^2\frac{d \Phi_{\text{DRAGON}}}{d E} + \frac{\langle\sigma v\rangle}{2}  \frac{d N}{d E}  \frac{\Delta \Omega\langle J\rangle_{\Delta \Omega}}{4 \pi m_{\mathrm{DM}}^{2}}
  \label{eq:function_fit} 
\end{equation}

To perform the fits, the expected annihilation gamma-ray signal $dN/dE$ given by \cite{Cirelli_2011, Ciafaloni2010} is convoluted to a Gaussian distribution (with $\sigma/E = 0.10$) to take into account the energy resolution of the H.E.S.S. telescope \cite{Montanari_2022buj}.

\begin{figure}[t!] 
  \centering 
    \includegraphics[width=7.5cm]{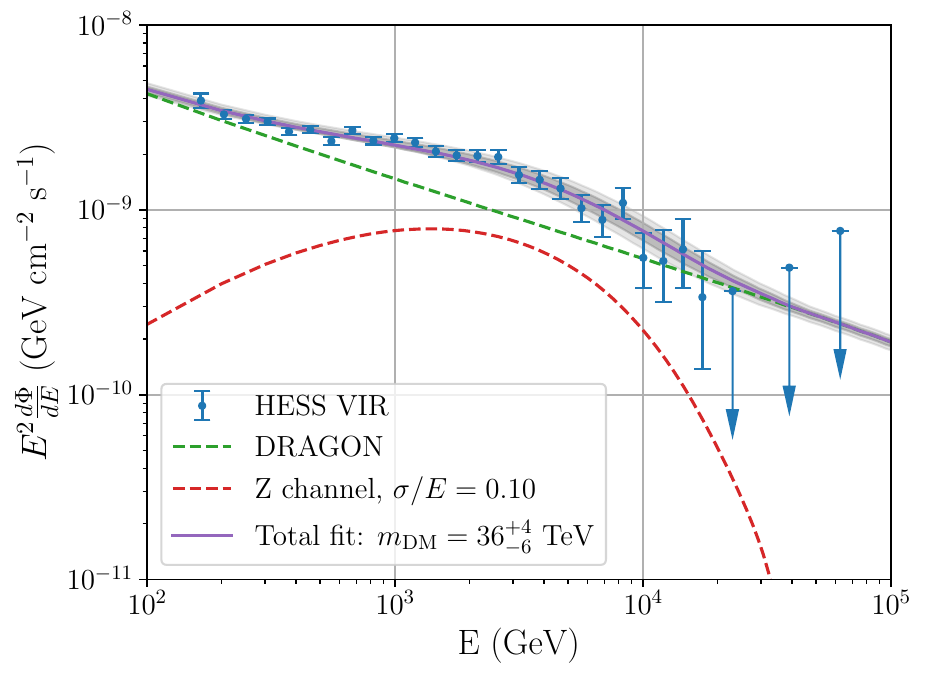}
    \includegraphics[width=7.5cm]{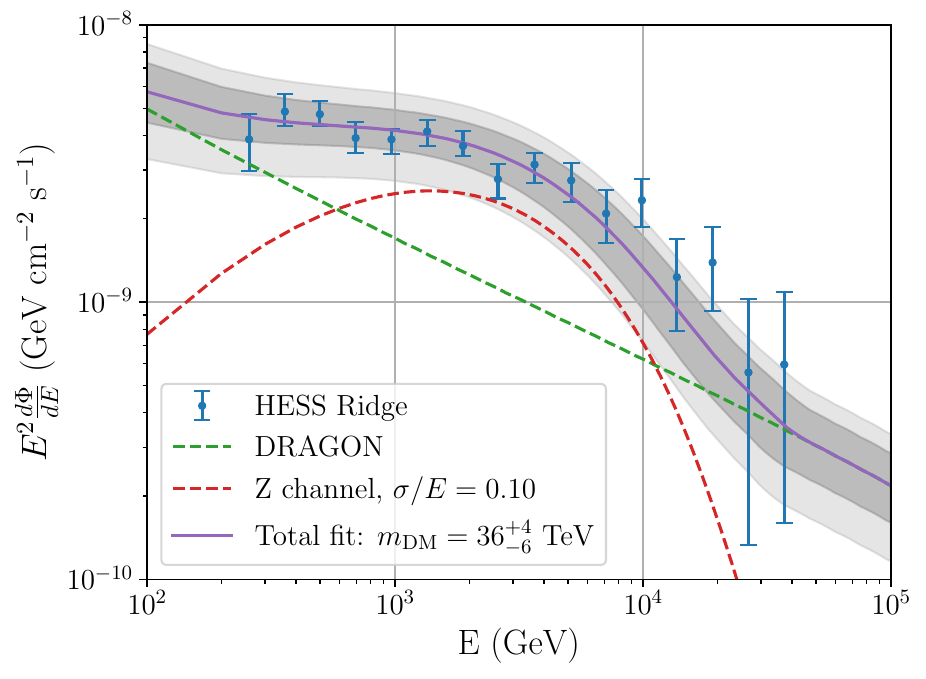}
    \caption{\footnotesize{Best fit (magenta line) to the H.E.S.S. data of the VIR (left panel) and Ridge (right panel). The background (green-dashed line) and DM component (red-dashed line) are also shown. The best-fit parameters (Table \ref{tab:VIR_Ridge_Diffuse}) show an agreement with a gamma-ray signal produced by the annihilation of a $36^{+4}_{-6}$ TeV WIMP particle and a background component, taking into account the different background component as modeled by DRAGON in the different regions (see text for further details). The dark grey and light gray regions correspond to the 1$\sigma$ and 2$\sigma$ uncertainty, respectively.} }
\label{fig:VIR_Ridge} 
\end{figure}

\begin{figure}[t!] 
  \centering 
  \includegraphics[width=7.5cm]{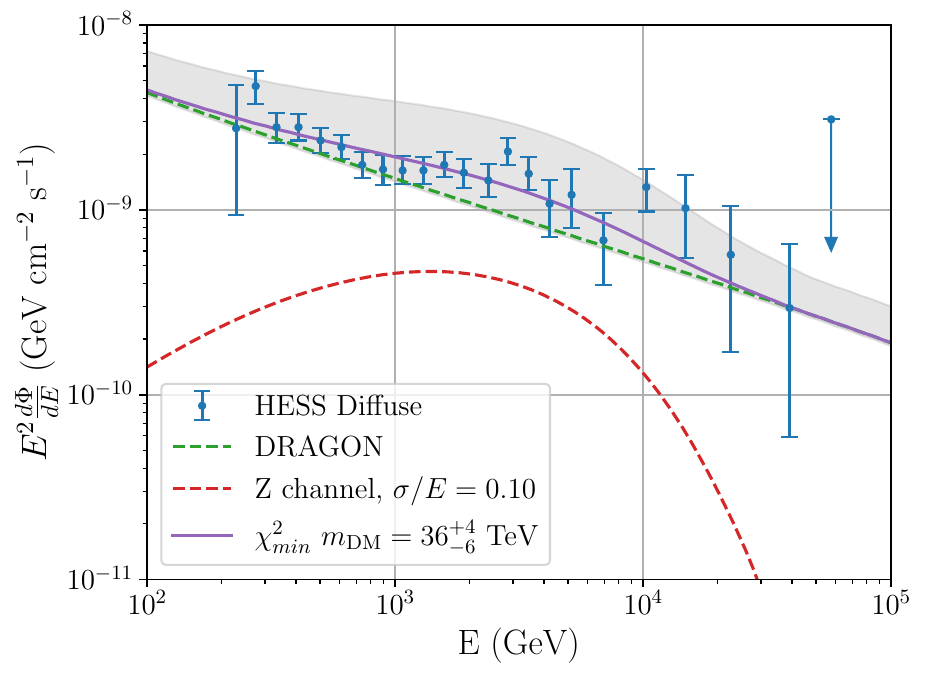}
    \includegraphics[width=7.5cm]{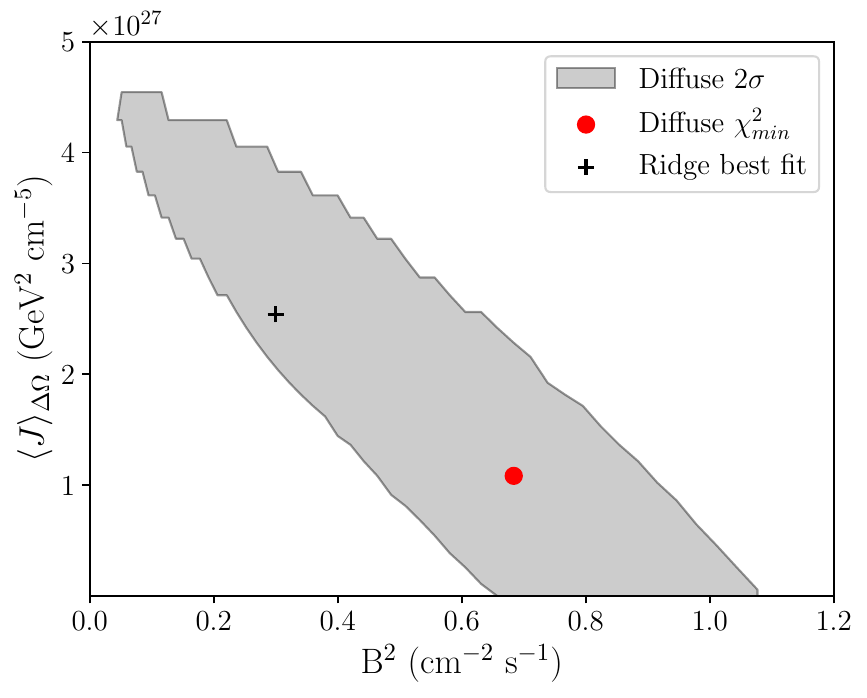} 
    \caption{\footnotesize{Left Panel: same as Figure \ref{fig:VIR_Ridge} for the Diffuse region. The light grey band is the 2$\sigma$ confidence level. The best-fit parameters are shown in Table \ref{tab:VIR_Ridge_Diffuse}. Right Panel: 2$\sigma$ confidence level region (in grey) for the J-factor $\langle J\rangle_{\Delta \Omega}$ and $B^2$ parameter space. The red dot is the best fit of the $\chi^2$ analysis, with $\chi ^2_{\text{min}}/ddof$ = 0.89. Note that the $2\sigma$ region is compatible with $\langle J\rangle_{\Delta \Omega} = 0$, indeed no DM signal. The best fit for the Ridge region (black cross) is compatible with the Diffuse best fit within the $2\sigma$ confidence level.}}
\label{fig:Diffuse} 
\end{figure}

\begin{table}[h!]
    \begin{center}
        \begin{tabular}{|c|c|c|c|}
        \hline
        \hline
Parameters                     & VIR                          & Ridge                       & Diffuse\\ 
\hline
\hline
$m_{\text{DM}}$ (TeV)                       & $36 ^{+7}_{-10}$ & -- & -- \\ \hline
$B^2$ ($\text{cm}^2 \text{s}^{-1}$)     
                                            & $9.2^{+0.8}_{-0.9}$                   & $0.3^{+0.2}_{-0.1}$                  & $0.8^{+0.2}_{-0.6}$  \\ \hline
$\langle J\rangle_{\Delta \Omega}$  ($ \text{GeV} ^2 \text{cm}^{-5}$)                             
                                            & $2.7^{+1.0}_{-0.6} \times 10 ^{28}$   & $2.5^{+1.0}_{-0.9} \times 10 ^{27}$  & $1.1^{+3.4}_{-1.1} \times 10 ^{27}$  \\ \hline
$\langle J\rangle_{\Delta \Omega} / J_{\text{DM-only}}$                                                              
                                            & $1000 ^{+400}_{+200}$                   & $1000 ^{+400}_{-400}$                 & $300^{+800}_{-300}$ \\ \hline
$\chi ^2$ / ddof                            & 0.99                                 & 1.04                                  & 0.89      \\ \hline
$\Delta \Omega $ (sr)                       & $9.57 \times 10^{-6}$                 & $3.26 \times 10^{-4}$                & $1.41 \times 10^{-4}$ \\
\hline
        \end{tabular}
        \caption{\footnotesize{Best-fit parameters for the VIR, Ridge and Diffuse regions (Figures \ref{fig:VIR_Ridge} and \ref{fig:Diffuse}). The DM mass $m_{\text{DM}}$ corresponds to the best fit in the VIR and it has been kept fixed in the rest of the regions. The $B$ parameter is the normalization of the DRAGON background model, and $\langle J\rangle_{\Delta \Omega}$ is the J-factor fitted. We also show the boost factor to the reference value for an NFW profile, i.e. $J_{\text{DM-only}}^{\text{VIR}} =  2.6 \times 10 ^{25} \text{GeV}^2 \text{cm}^{-5}$, $J_{\text{DM-only}}^{\text{Ridge}} =  2.5 \times 10 ^{24} \text{GeV}^2 \text{cm}^{-5}$ and $J_{\text{DM-only}}^{\text{Diff}} =  4.2 \times 10 ^{24} \text{GeV}^2 \text{cm}^{-5}$. Finally, the $\chi^2 / ddof$ of each fit is shown with the solid angle $\Delta \Omega$ of each region. The uncertainties of the values correspond to a $2\sigma$ confidence level.}}
        \label{tab:VIR_Ridge_Diffuse}
    \end{center}
\end{table}

\subsubsection*{VIR}

In Figure \ref{fig:VIR_Ridge} (left panel) we show the fit of the DM plus DRAGON background component (Equation \ref{eq:dflux/de-total}) to the H.E.S.S. data in the VIR. Following previous analyses \cite{Cembranos_2013, 2012PhRvD..86j3506C}, the inclusion of a background component in the analysis of the VIR region is required in order to well fit the low energy H.E.S.S. data, and in agreement with high energy Fermi-LAT data, whose angular resolution above $\sim10$ GeV is $\theta\sim 0.1^\circ$, comparable to H.E.S.S.. Although we do not include Fermi-LAT data in our analysis our background model DRAGON and renormalization factor $B^2$ are in agreement with the gamma-ray spectra detected by Fermi-LAT at energy $\text{E} > 10$ GeV (see in the Conclusions Figure \ref{fig:best_profiles}, upper left panel). The fitted parameters, i.e. the DM mass $m_{\text{DM}}$, background renormalization $B$ and astrophysical J-factor $\langle J\rangle_{\Delta \Omega}$ are shown in Table \ref{tab:VIR_Ridge_Diffuse}: the good quality of the fits show an agreement of the data with the hypothesis of a DM component to the gamma-ray signal, produced by a $m_{\text{DM}} =  36^{+4}_{-6}$ TeV DM particle annihilating in the ZZ channel at GC, with a J-factor $\langle J\rangle_{\Delta \Omega} = 2.7^{+0.6}_{-0.3} \times 10 ^{28}$ $\text{GeV}^2 \text{cm}^{-5}$. This result is compatible with previous works\cite{Cembranos_2013, 2012PhRvD..86j3506C}. Minor differences may be explained by: 1) newly updated data by the H.E.S.S. collaboration in the region \cite{HESSI, HESSII}; 2) the inclusion of the DRAGON background model instead of the simple power-law \cite{2012PhRvD..86j3506C, Cembranos_2013}; 3) the use of a different Monte Carlo event generator software to model the gamma-ray annihilation flux \cite{2013JHEP...09..077C}. In Table \ref{tab:VIR_Ridge_Diffuse} we also give the $\langle J\rangle_{\Delta \Omega} / J_{\text{DM-only}}$ es reference value: this value represents the enhancement factor (when compared to a benchmark NFW profile) required to the DM density distribution to be the origin of such a gamma-ray flux. This boost factor represents the first hint to assume the existence of a more cuspy profile or a DM spike, as we discuss in the next Section \ref{J_factors_morphology}. Following this idea, it is straightforward to expect that the DM signal could manifest only locally over the background, remaining undetected in outer regions. Indeed, we keep the best fit of the DM mass obtained in this region to develop the fits in the Ridge, Diffuse, Halo and IGS, where the DM component is expected to be subdominant.

\subsubsection*{Ridge}

In Figure \ref{fig:VIR_Ridge} (right panel) we show the data observed by H.E.S.S. in the Ridge, fitted to the Equation \ref{eq:dflux/de-total}. Here, we fixed the DM mass to the best-fit value obtained in the analysis of the VIR region, and we fit here only two free parameters: the background renormalization $B$ and the astrophysical J-factor $\langle J\rangle_{\Delta \Omega}$. The best-fit parameters are given in Table \ref{tab:VIR_Ridge_Diffuse}. We have also verified that without fixing the $m_{\text{DM}}$ to the VIR value, we get a similar result with higher uncertainty.

\subsubsection*{Diffuse}
\label{sec:diffuse}

As for the VIR and the Ridge regions, we perform a $\chi^2$ analysis (Equation \ref{eq:chi_2_formula}) of the Diffuse region data. In Figure \ref{fig:Diffuse} (right panel) we show the best-fit parameters (red dot) and the 2$\sigma$ confidence level (the grey region in the same figure). The best-fit parameters are also shown in Table \ref{tab:VIR_Ridge_Diffuse}. The 2$\sigma$ region is compatible with $\langle J\rangle_{\Delta \Omega}$ = 0, meaning that the DM gamma-ray component is subdominant to the background signal in this region, and the result could be considered as an upper limit to the J-factor. This result is also compatible with the hypothesis that most of the DM-related gamma-ray signature is hidden in this region, being the contribution of the VIR covered by the applied masked.

\subsubsection*{Halo and IGS}

\begin{figure}[t!] 
  \centering 
   \includegraphics[width=0.49\textwidth]{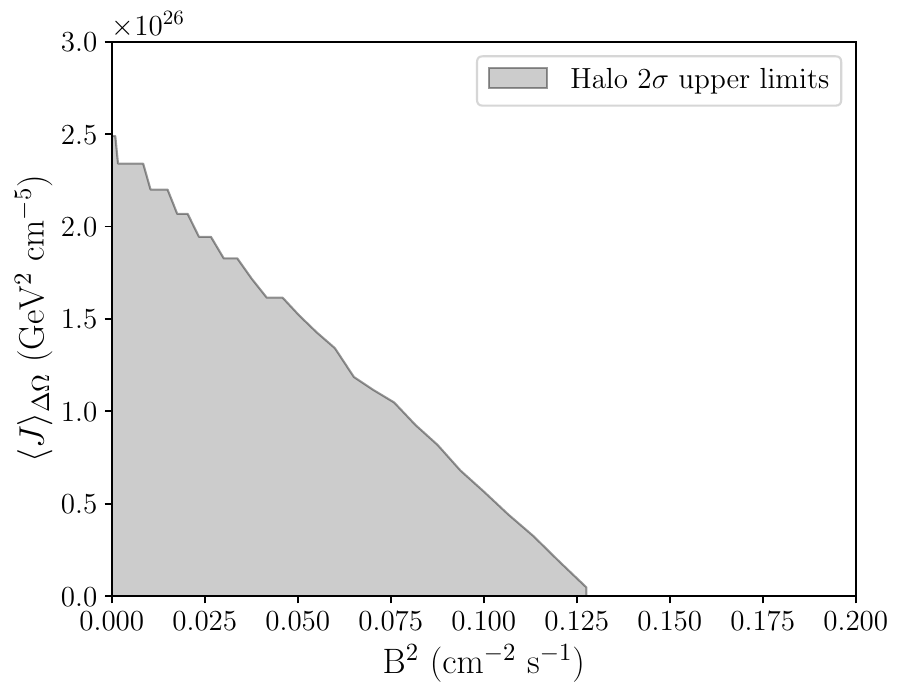}  
    \includegraphics[width=0.49\textwidth]{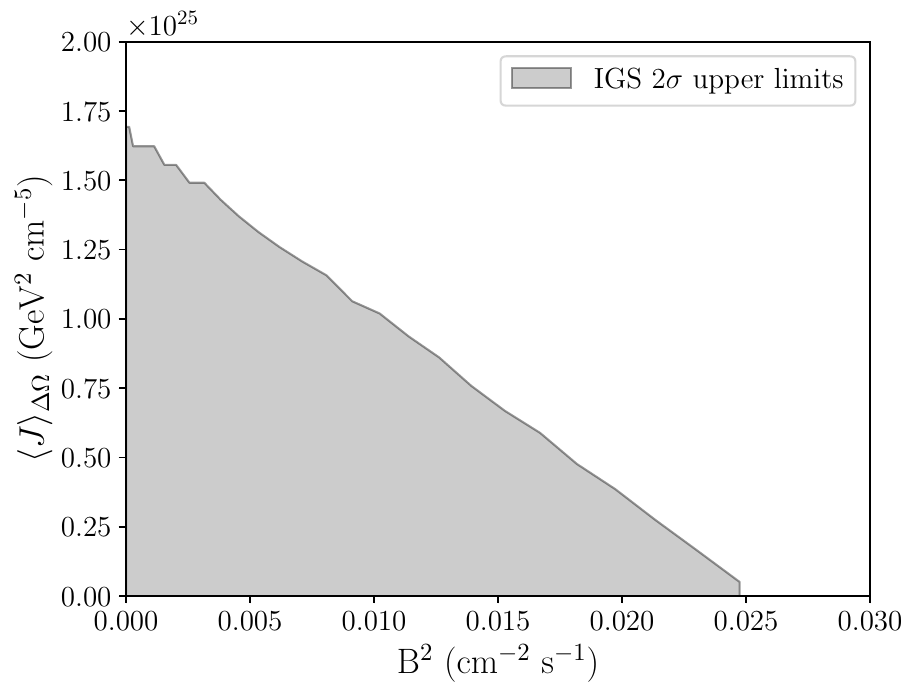} 
    \caption{\footnotesize{2$\sigma$ upper limits on the J-Factor obtained by the analysis of the Halo region (left panel) and IGS region (right panel). The $B^2$ parameter corresponds to the renormalization of the DRAGON background model and the grey area corresponds to the 2$\sigma$ upper limits. The upper limits are computed assuming the thermal relic cross-section $\langle\sigma v\rangle = 2.2 \times 10^{-26} \text{cm}^3\text{s}^{-1}$ and with a DM particle of $m_{\text{DM}} = 36^{+4}_{-6}$ TeV, the best-fit value in the VIR, annihilating in the ZZ channel.}}
\label{fig:Halo_IGS} 
\end{figure}

\begin{table}[t!]
    \begin{center}
        \begin{tabular}{|c|c|c|}
\hline
\hline 
\textbf{Parameters}                              & Halo             & IGS\\ \hline 
\hline
$m_{\text{DM}}$ (TeV)                            & -- & --    \\ \hline
$B^2_{\text{UL}}$ ($\text{cm}^2 \text{s}^{-1}$)
                                                 & 0.13                      & 0.02          \\ \hline
$\langle J\rangle_{\Delta \Omega}^{\text{UL}}$  ($\text{GeV}^2 \text{cm} ^{-5}$) 
                                                 & $2.5\times 10^{26}$      & $1.7\times 10^{25}$     \\ \hline
$\Delta \Omega$ (sr)                             & $5.97 \times 10^{-4}$     & $6.38 \times 10^{-3}$  \\ \hline 
        \end{tabular}
        \caption{\footnotesize{Upper limits within 2$\sigma$ confidence level on the $B^2$ and J-factor parameters for the Halo and IGS regions, calculated in the respective solid angle $\Delta \Omega$. The value of the DM particle mass is fixed, in agreement with the VIR best-fit value, i.e. 36 TeV. The reported values are the maximum values allowed within the grey area in Figure \ref{fig:Halo_IGS}. For comparison, $J_{\text{DM-only}}^{\text{Halo}} =  1.8 \times 10 ^{24} \text{GeV}^2 \text{cm}^{-5}$ and $J_{\text{DM-only}}^{\text{IGS}} =  3.6 \times 10 ^{21} \text{GeV}^2 \text{cm}^{-5}$.}}
        \label{tab:Diff_Halo_IGS}
    \end{center}
\end{table}

In the Halo \cite{HESSHalo, PhysRevLett.106.161301} and IGS \cite{PhysRevLett.129.111101} regions, no signal is detected by the H.E.S.S. collaboration. In both regions, the flux in the ON region is compatible with the flux in the OFF control region, adopted for background rejection. Indeed, in this case, we adopt a different procedure in order to set upper limits on the DM density distribution and astrophysical J-factor, and we ask for a 2$\sigma$ detection of the theoretical gamma-ray flux, i.e. $\Phi_{\text{th}} = \Phi_{\text{DM}} + \Phi_{\text{DRAGON}}$. We determine the background given by DRAGON for the same regions and masks, and we scan the J-factor and $B$ parameter space for a thermal WIMP mass of $36$ TeV annihilating in the ZZ channel, consistently with the fit in the VIR, Ridge and Diffuse (see Figure \ref{fig:Halo_IGS} and Table \ref{tab:Diff_Halo_IGS}). Indeed, we have  \cite{2018PhRvD..98h3008G, 1983ApJ...272..317L}:

\begin{equation}
  \frac{\Phi_{\text{th}} \sqrt{A_{\text{eff}} t_{\text{exp}} \Delta \Omega}}    {\sqrt{\Phi_{\text{HESS}} + \Phi_{\text{th}}}} < 2,
  \label{eq:xi_Halo_IGS} 
\end{equation}

where $A_{\text{eff}}$ is the H.E.S.S. effective area, $t_{\text{exp}}$ is the exposure time, $\Delta\Omega$ is the solid angle and $\Phi_{\text{HESS}}$ is the integrated flux observed by H.E.S.S., compatible with the background component.

\subsection{Results}
\label{fits_Jfactors}

So far, we have determined the upper limits on the J-factor for the prospective thermal multi-TeV DM candidate of $m_\text{DM}=36^{+4}_{-6}$ TeV in 5 concentric regions of the GC observed by H.E.S.S. These J-factors have been determined as the amplitude required for the DM component to fit the H.E.S.S. data. It is well known that the J-factor is, by definition, the integral of the squared DM density distribution profile along the line of sight (Equation \ref{eq:astrophysicalfactor}). Indeed, within our hypothesis, we can set upper limits on the radial DM density distribution profile in the Galaxy by the study of the gamma-ray spectra. 

\begin{figure}[t!] 
  \centering 
  \includegraphics[width=0.7\textwidth]{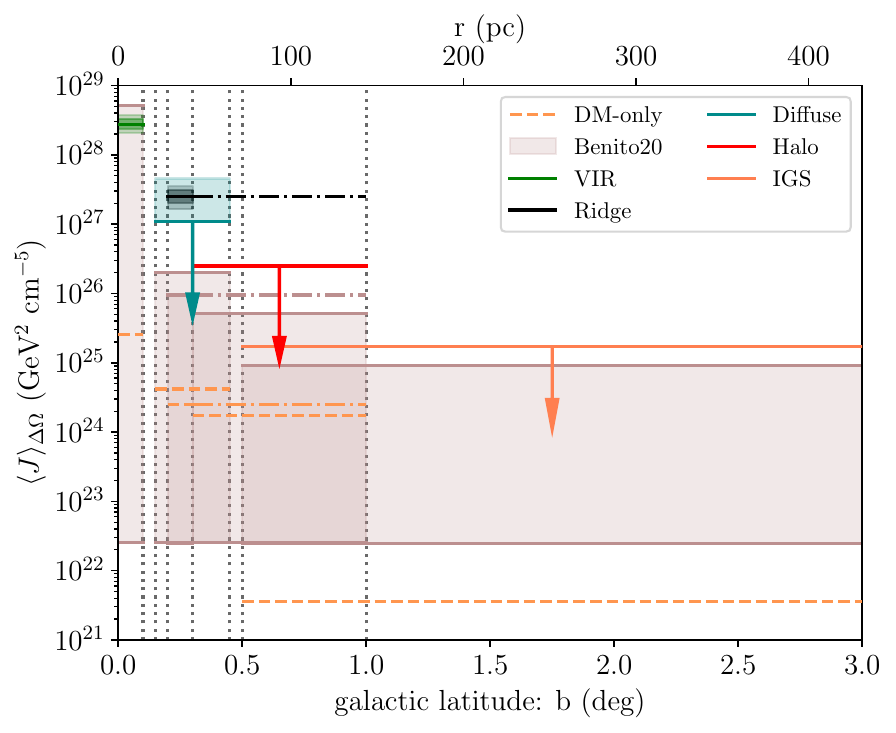} 
  \caption{\footnotesize{J-factor $\langle J\rangle_{\Delta \Omega}$ obtained by the study of the gamma-ray spectra in 5 different regions of the GC, where $r$ is the distance from the GC in pc and b represents the galactic latitude (deg). Each region is represented by the vertical grey dotted lines, corresponding to their boundaries. For the VIR and Ridge regions, the best fit and the 1$\sigma$ and 2$\sigma$ uncertainties are represented; for the Diffuse region, the best fit and the 2$\sigma$ uncertainty are shown, where the lower limit of the 2$\sigma$ confidence level is compatible with $\langle J\rangle_{\Delta \Omega} = 0 $, as explained in the right panel of Figure \ref{fig:Diffuse}. Finally, for the Halo and IGS regions, the values correspond to the 2$\sigma$ upper limit. For comparison, we show the J-factors computed with the DM-only profile (orange dashed lines) and with the Benito20 DM profile \cite{Benito20} (brown area). Here, the DM mass is $m_{\text{DM}} = 36^{+4}_{-6}$ TeV and we assume a thermal annihilation cross section $\langle\sigma v\rangle \simeq 2.2\times 10 ^{-26} \text{cm}^3 \text{s}^{-1}$. Due to the non-spherical symmetry of the Ridge region, we show with the dot-dashed lines the extension of the region in the galactic longitude, up to $1^\circ$.}}
  \label{fig:J_factor_fit_vs_EVANS} 
\end{figure}

In Figure \ref{fig:J_factor_fit_vs_EVANS} we show the J-factors obtained by the analysis of the gamma-ray spectra. We compare our results with the J-factors obtained for the same regions by assuming the benchmark DM-only NFW profile (orange dashed lines) and the Benito20 model (brown region). From the analysis of the VIR, we find out J-factors higher than the benchmark NFW profile, in agreement with the range of J-factors obtained by integrating the DM profile obtained in Benito20, extrapolated up to those regions. In particular, a boost factor of $\sim 1000$ over the benchmark NFW profile is required to explain the fitted values. In the Diffuse and Ridge regions, the best fit of the gamma-ray spectra favors a DM distribution profile steeper than the extrapolation of the Benito20's profile across the GC. Note that, for the Diffuse, as explained in the right panel of Figure \ref{fig:Diffuse}, the lower limit of the 2$\sigma$ confidence level is compatible with $\langle J\rangle_{\Delta \Omega} = 0 $ and, therefore, the results are shown as upper limits. Finally, in the Halo and IGS outer regions the spectral upper limits are not that restrictive.

It is worth mentioning that the dynamical data studied in Benito20 is able to give constraints on the DM distribution only to $\text{r} \gtrsim 2.5$ kpc from the GC. Indeed, our study represents a complementary analysis to solve the issue of the DM distribution in the Galaxy. Our results are the first upper limits on the DM distribution for the innermost Galactic region within $\text{r} \lesssim 400$ pc, in agreement with the previous constraints on the DM distribution in the outer region of the Galaxy. In the following section, we study the possibility to disentangle within our study a cuspy DM profile from a DM spike. In fact, the enhancement of the DM density distribution in the VIR (when compared to the benchmark NFW profile) could be explained by the interaction of the WIMP particles with both the gravitational potential of the SMBH Sgr A* and the baryonic component of the region.

\section{The innermost dark matter density profile}
\label{J_factors_morphology}

\begin{figure}[t!] 
  \centering 
  \includegraphics[width=9cm]{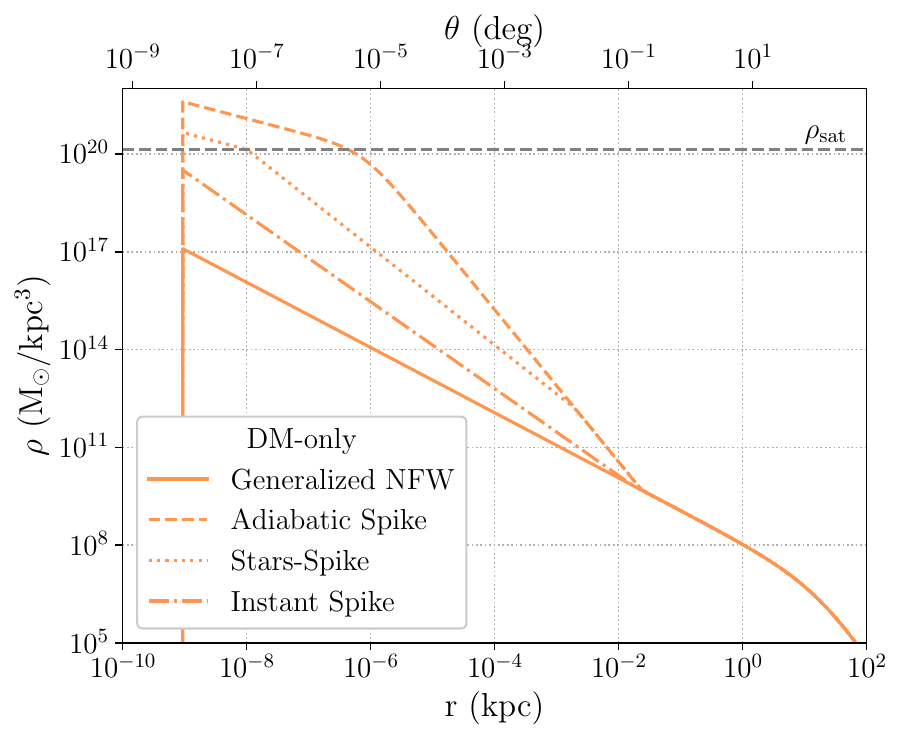} 
  \caption{\footnotesize{The 4 different models of the DM distribution in the innermost region of the GC applied to the DM-only profile: extrapolation of the underlying generalized NFW profile to the innermost GC region (solid line), adiabatic spike (dashed line), stars-spike (dotted line), and instant spike (dotted-dashed line). The grey horizontal line represents the annihilation plateau for a mass of $m_{\text{DM}} = 36$ TeV (see text for details).}}
  \label{fig:DMOnly_all} 
\end{figure}

In this section, we compare our results with several DM density distribution profiles, obtained by both cosmological simulations and studies of Galactic dynamics. First, we consider the generalized NFW profile (Equation \ref{eq:NFW_general}), by extrapolating the profiles to the innermost GC region without any modifications to the inner part. Second, we investigate the possibility that the DM profile could be modified in the central region due to the presence of the SMBH Sgr A*. We consider different models: 1) a DM spike formed adiabatically \cite{PhysRevLett.83.1719, Sadeghian2013}; 2) the effect of the interaction with the stars of the bulge with the adiabatic spike \cite{PhysRevD.78.083506, doi:10.1142/S0217732305017391}; 3) a DM spike formed instantaneously \cite{Ullio2001}; and 4) we review the case for the rotating Kerr BH. As an example, in Figure \ref{fig:DMOnly_all} we show all those models applied to the benchmark NFW profile. Furthermore, it is well known that the DM density distribution in the innermost GC region cannot increase indefinitely \cite{PhysRevLett.83.1719}. Instead, in the WIMP framework, it is typically expected to reach a maximum plateau due to the growth of the annihilation rate with the enhancement of the DM density itself. The maximum value of the annihilation plateau depends on the WIMP mass. For a WIMP particle of $m_{\text{DM}} = 36$ TeV the central plateau is $\rho_{\text{sat}} = m_{\text{DM}} / (t_{\text{BH}} \langle\sigma v\rangle) = 1.4\times 10^{20}$ $\text{M}_{\odot} \text{kpc}^{-3}$, taking an assumed age for the SMBH Sgr A* $t_{\text{BH}} = 10^{10}$ yr \cite{Lacroix_2018, Gammaldi_2016}. This plateau is also shown in Figure \ref{fig:DMOnly_all}.
In the following sections, we will discuss in detail all these models and how they affect the J-factor in the different regions of the GC analyzed so far. 

\begin{figure}[t] 
  \centering 
  \includegraphics[width=0.49\textwidth]{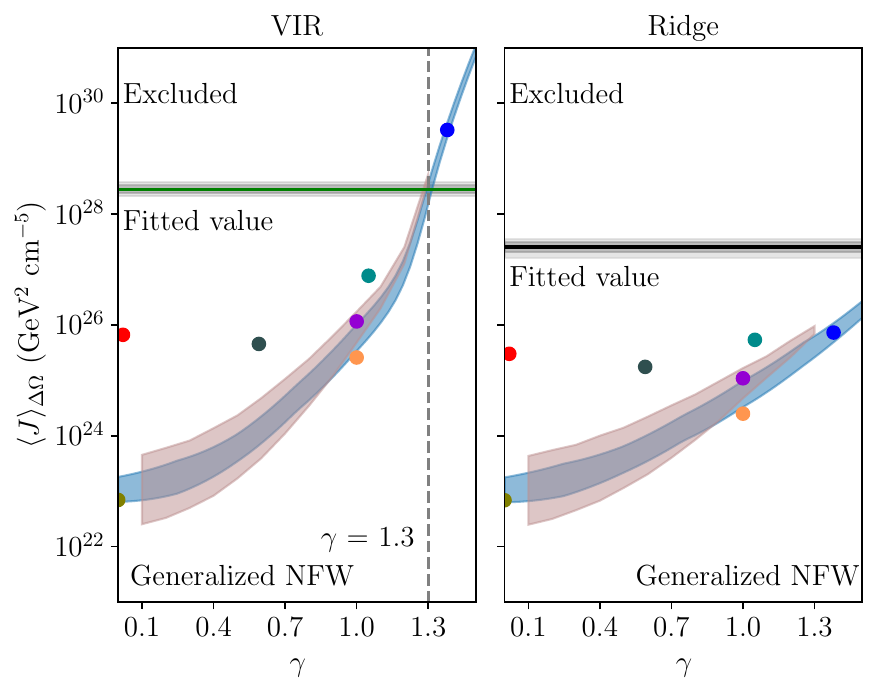} 
  \includegraphics[width=0.49\textwidth]{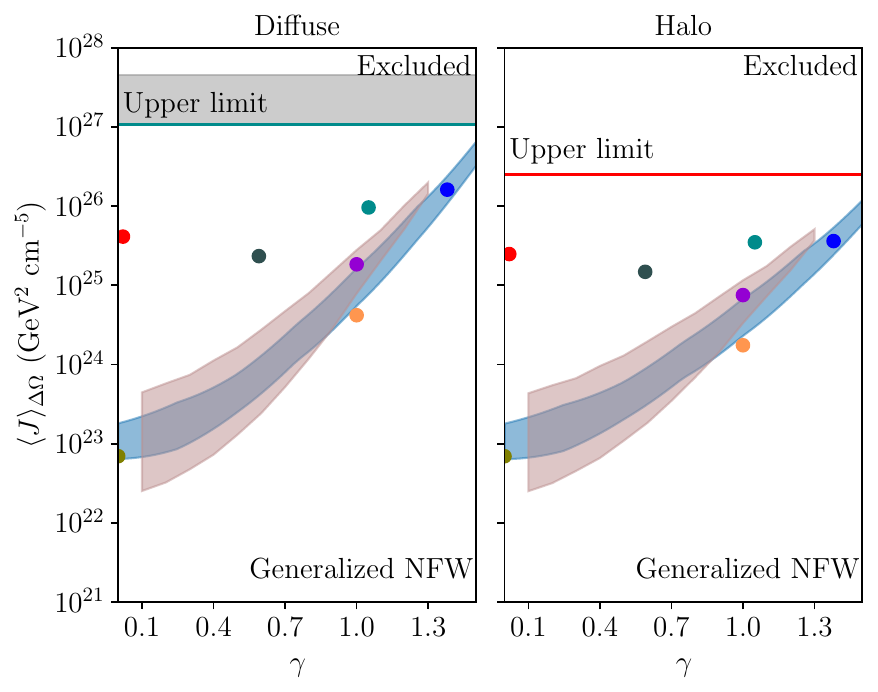} 
  
  \includegraphics[width=0.49\textwidth]{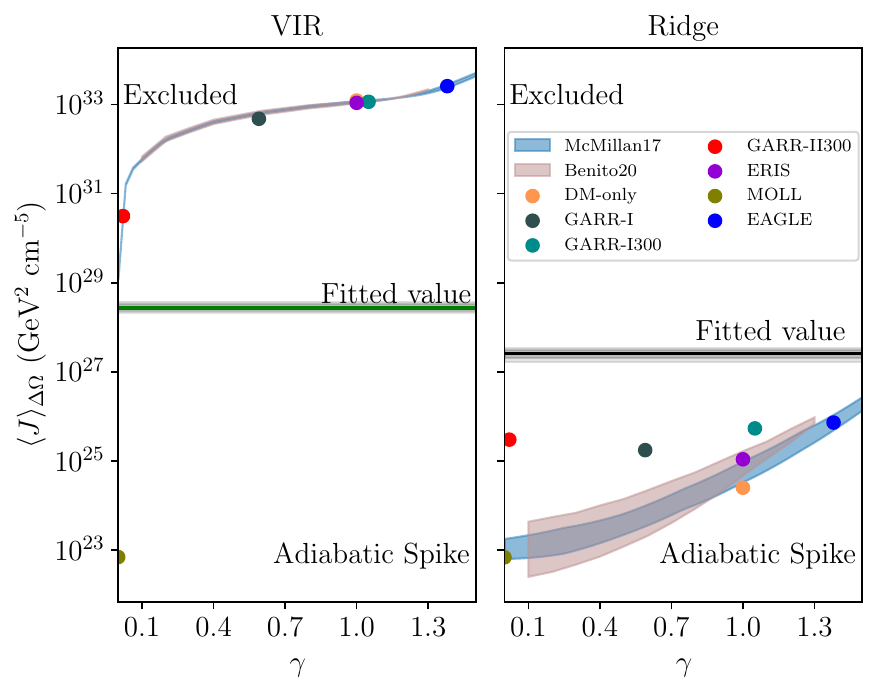} 
  \includegraphics[width=0.49\textwidth]{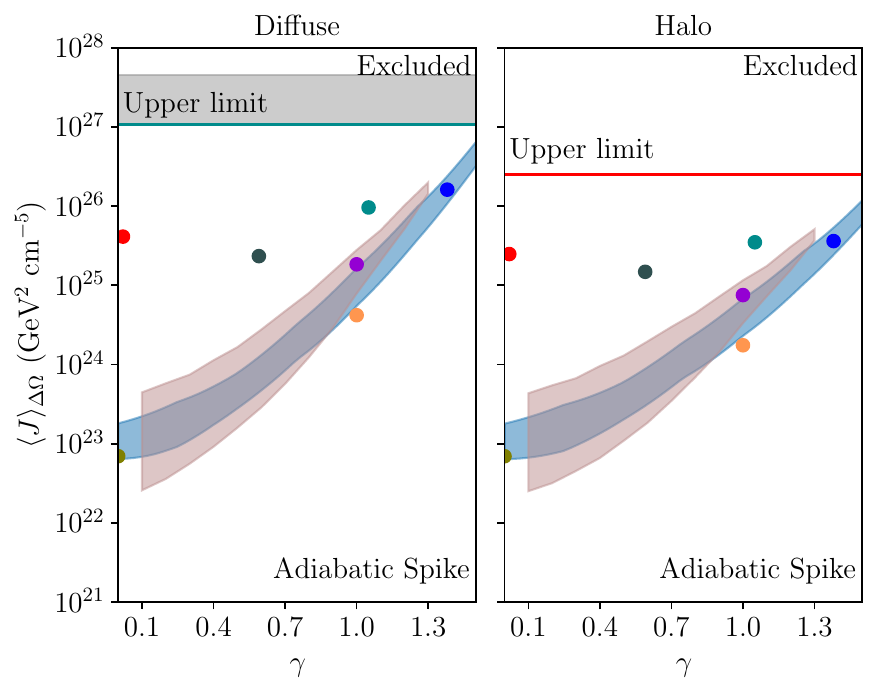}

  \caption{\footnotesize{J-factor $\langle J\rangle_{\Delta \Omega}$ as a function of the density slope $\gamma$, for the approaches: generalized NFW (first row) and DM adiabatic spike (second). We show, for each row the regions VIR (first panel from left to right), Ridge (second panel), Diffuse (third panel) and Halo (fourth panel). The horizontal lines are the fitted values from the gamma-ray spectra, assuming the thermal relic cross-section $\langle\sigma v\rangle \simeq 2.2\times 10 ^{-26} \text{cm}^3 \text{s}^{-1}$ and a fitted DM mass $m_{\text{DM}} = 36^{+4}_{-6}$ TeV, with their uncertainties represented in grey: $1\sigma$ and $2\sigma$ for the VIR and Ridge, 2$\sigma$ for the Diffuse and the Halo 2$\sigma$ upper limit. The colored dots correspond to DM density models from N-body simulations and the colored regions to the observational models: McMillan17 \cite{McMillan2017} in blue and Benito20 \cite{Benito20} in brown. The vertical gray dashed line indicates when the integrated J-factors cross the fitted values. Above them, the predicted gamma-ray flux exceeds the observed flux by H.E.S.S. and, therefore, this part of the parameter space is excluded. }
  } 
  \label{fig:JFactors_NFW_Spike} 
\end{figure}

\subsection{Generalized NFW profile}
\label{J_factors_NFW}

In Figure \ref{fig:JFactors_NFW_Spike} (upper row) we compare the J-factors obtained by the spectral analysis with the J-factors expected by both cosmological simulations (dots of different colors) and dynamical constraints (blue and brown regions). Here, the outer profiles are extrapolated up to the innermost region to determine the expected value of J-factors in the different regions. We only show the more restrictive regions: VIR and Ridge (first and second panel) and Diffuse (third panel). As a consistency check, we also show the Halo region (fourth panel). We do not show the results for the IGS region, which is the less constraining region. 

To explain the gamma-ray flux observed in the VIR as a DM annihilation signal we would need a generalized NFW  with $\gamma \sim 1.3$. This result is consistent with the range of parameters allowed by the dynamical constraints of McMillan17 and Benito20. Interestingly, this result is also in agreement with the one obtained for the Galactic Center Excess observed by Fermi-LAT in the GeV range ($\gamma =  1.1$-$1.2 $, \cite{Di_Mauro_2021}), although the WIMP candidate should indeed have a different mass. Nonetheless, the comparison of our results with the expected J-factors in the Ridge, Diffuse and Halo regions does not allow us to exclude any profile. This is consistent with the hypothesis that the DM signal is subdominant in these regions, as the DM distribution is less dense, lacking the enhancement factor which could allow for the detection of the DM signature in the innermost part of the Galaxy. To conclude, with this first analysis we can rule out any DM distribution profile which could produce a gamma-ray flux above the H.E.S.S. data. Indeed, we can exclude generalized NFW profiles with $\gamma \gtrsim 1.3$.

\subsection{Adiabatic Dark matter Spike}
\label{J_factors_Spike_Stars}

The existence of the SMBH Sgr A* in the GC with a mass $M_{\text{BH}} = 4.297 \pm 0.012 \times10^6 \text{M}_{\odot}$ \cite{GRAVITY_2021} has been recently demonstrated \cite{akiyama2022first}. The possibility to have an enhancement of the DM density due to the gravitational interaction with the SMBH Sgr A* has been largely studied in the literature. Among other models, a DM spike is considered to be formed adiabatically if the SMBH has grown in the center of the Galaxy, very slowly compared to the typical dynamical timescales of the region, and without any occurrence of big mergers in the Milky Way for the last $\sim 10$ Gyr \cite{2211.01006} (a conservative assumed age for the SMBH Sgr A* \cite{Lacroix_2018}). 

\begin{center}
\begin{table}[t!]
\begin{center}
\resizebox{\textwidth}{!}
{
\begin{tabular}[b]{|c|c|c|c|c|c|c|c|}
\hline
\hline

Profile & $\gamma$ & $\gamma_{\text{sp}}$ & $R_{\text{sp}}$ \text{(pc)} & $\theta_{\text{sp}}$ \text{(deg)} & $\gamma_{\text{inst}}$ & $R_{\text{inst}}$ \text{(pc)} & $\theta_{\text{inst}}$ \text{(deg)} \\
\hline
\hline

DM-only & 1 & $2.33$ & $23.5$ & $0.17$ & $1.33$ & $16.9$ & $0.12$ \\
\hline
GARR-I & 0.59 &$2.29$ & $16.1$ & $0.11$ & $1.23$ & $13.2$ & $0.09$ \\
\hline
GARR-I300 & 1.05 & $2.34$ & $10.2$ & $0.07$ & $1.35$ & $8.0$ & $0.06$ \\
\hline
GARR-II300 & 0.02 &$2.25$ & $2.8$ & $0.02$ & $1.14$ & $19.0$ & $0.13$ \\
\hline
ERIS & 1 &$2.33$ & $16.2$ & $0.1$ & $1.33$ & $11.6$ & $0.08$ \\
\hline
MOLL & $8\times 10^{-9}$ & $2.25$ & $0.03$ & $0.0002$ & $1.12$ & $88.0$  & $0.62$ \\
\hline
EAGLE & 1.38 &$2.38$ & $6.8$& $0.05$ & $1.48$ & $72.6$ & $0.51$ \\
\hline
McMillan17 & 0 - 1.5 & $2.25$ - $2.40$ & $3.8$ - $47.6$ & $0.03$ - $0.33$ & $1.12$ - $1.41$ & $80.5$ - $11.7$* & $0.56$ - $0.08$*\\
\hline
Benito20 & 0.1 - 1.3 & $2.26$ - $2.37$ & $6.9$ - $61.3$ & $0.05$ - $0.43$ & $1.14$ - $1.44$ & $98.4$ - $10.1$* & $0.70$ - $0.07$* \\
\hline

\end{tabular}
}
\end{center}
\caption{\footnotesize{Parameters used for the adiabatic spike (Equation \ref{eq:DM_profile_spike}) and instantaneous spike (Equation \ref{eq:DM_profile_instant}) for different DM density profiles: DM-only simulation, the Hydrodynamical simulations GARR, GARR-I, GARR-I300, GARR-II300, ERIS, MOLL and EAGLE (see \cite{Gammaldi_2016} and the references within) and the range of values given by the observational models McMillan17 \cite{McMillan2017} and Benito20 \cite{Benito20}. * Larger $R_{\text{inst}}$ corresponds to smaller $\gamma$.}}
\label{tab:spike_parameter_table}
\end{table}
\end{center}

The DM adiabatic spike density profile \cite{PhysRevLett.83.1719, Sadeghian2013} can be described with the modification of the inner part as $\rho_{\text{sp}}(r) \simeq \rho_{\text{R}}(1-\frac{2 R_{\text{s}}}{r})^3 (\frac{r}{R_{\text{sp}}})^{-\gamma_{\text{sp}}}$. In this description, both the slope and the radius of the spike depend on the inner slope $\gamma$ of the underlying DM density profile: the spike slope is  $\gamma_{\text{sp}} =  (9 - 2 \gamma)/(4 - \gamma)$ and the spike extends up to radius $R_{\text{sp}} = \alpha_{\gamma} r_{s} (M_{\text{BH}}/ (\rho_s r_{s}^3))^{(1/(3- \gamma))}$; where $\alpha_{\gamma}$ is computed numerically and $\rho_{\text{R}} = \rho_{\text{halo}}(R_{\text{sp}})$ \cite{PhysRevLett.83.1719}. In Table \ref{tab:spike_parameter_table} we show the $\gamma_{\text{sp}}$ and $R_{\text{sp}}$ for all the DM profiles considered in this work. Regarding DM density spikes with DM self-annihilating particles, it is important to stress that the DM density can only reach a maximum value $\rho_{\text{sat}}$. This modifies the total DM density profile as follows:

\begin{equation}
  \rho (r) = 
     \begin{cases}
       0 & r < 2 R_{\text{S}}\\
       \frac{\rho_{\text{sp}}(r) \rho_{\text{sat}}(r)}{\rho_{\text{sp}}(r) + \rho_{\text{sat}}(r)}  &  2 R_{\text{S}} \leq r \leq R_{\text{sp}}\\ 
       \rho_{\text{halo}} (r) & r \geq R_{\text{sp}}\\
     \end{cases}
  \label{eq:DM_profile_spike}
\end{equation}

where $R_{\text{S}}$ is the Schwarzschild radius and $\rho_{\text{halo}}(r)$ is the underlying DM density distribution profile. In Figure \ref{fig:DMOnly_all} we show the adiabatic spike (dashed line) associated with the benchmark NFW profile (solid line), as an example. In the literature, the saturation factor $\rho_{\text{sat}}$ is usually taken as a central plateau $\rho_{\text{sat}} = m_{\text{DM}} / (t_{\text{BH}} \langle\sigma v\rangle)$, where $t_{\text{BH}}$ is the age of the central SMBH, with a value $\sim 10^{10}$ yr \cite{Lacroix_2018, Gammaldi_2016}. Here, we relax this plateau with the expression $\rho_{\text{sat}}(r) = m_{\text{DM}} / (t_{\text{BH}} \langle\sigma v\rangle) (\frac{r}{R_{\text{sat}}})^{-1/2}$, where this $-1/2$ slope appears as a correction when taking into account anisotropies in the velocity distribution of the DM particles in the GC \cite{Vasiliev2007}, being $R_{\text{sat}}$ the radius at which $\rho(R_{\text{sat}}) = \rho_{\text{sat}}$. Furthermore, we follow the relativistic approach \cite{Sadeghian2013} and assume that the DM density distribution vanishes at $r < 2R_{\text{S}}$ instead of $r < 4R_{\text{S}}$ as in the Newtonian analysis \cite{PhysRevLett.83.1719}.\\

In Figure \ref{fig:JFactors_NFW_Spike} (lower row), we compare the J-factor obtained by the spectral analysis with the J-factor expected in the case of having an adiabatic DM spike on different underlying DM profiles. We find out that, by analyzing the VIR region (first panel), we can exclude all profiles except very shallow ones like MOLL. Nonetheless, the latter enhancement would be indistinguishable in the gamma-ray spectra. Indeed, it would be compatible only with an astrophysical origin of the VIR spectral feature. The Ridge, Diffuse and Halo regions (rest of the panels) do not allow us to exclude any profile. In fact, all those 3 regions do not include the inner region inside $\theta \lesssim 0.15^\circ$ (Figure \ref{fig:regions_GC}). A negligible contribution of the prospective DM spike to the Diffuse region is in agreement with a DM spike angular extension smaller or compatible with the inner and outer radius observed by H.E.S.S., i.e. $\theta_\text{sp}\lesssim\theta_\text{Diff} = 0.15^\circ$-$0.45^\circ$ or $R_\text{sp} \lesssim R_\text{Diff} = 22$-$65$ pc, also in agreement with the result obtained by the spectral analysis in Section \ref{sec:diffuse}.

\subsubsection{Effect of stars on the spike}

The previous approach only takes into account the consequences of the inner DM density profile when considering the adiabatic growth of the SMBH, without considering any other interactions with other elements of the galaxy. As a correction, we can study the same case with the addition of the interactions with the stars of the bulge \cite{PhysRevD.78.083506, doi:10.1142/S0217732305017391}. In particular, with this approach that we name stars-spike in this paper, we are including the scattering of DM particles off of stars, which causes the dynamical heating of DM particles and, consequently, flattens the existing spike. Another process is the DM capture by the stars, but this process is secondary unless the cross-section of WIMP-baryons is sufficiently large. In summary, when taking into account the effect of the stars, inner profiles are less cuspy than the adiabatic spike case \cite{PhysRevD.78.083506, doi:10.1142/S0217732305017391}, lowering the internal slope to $\gamma_{\text{star}} = 1.5$ inside the radius of influence of Sgr A*, i.e. $r_{\text{b}} = 2$ pc. In Figure \ref{fig:DMOnly_all} we show the change of the profile (dotted line) with respect to the adiabatic case (dashed line). 

As a consequence, the values of the J-factors are also generally smaller (upper row of Figure \ref{fig:JFactors_stars_instant}). In this case, for the VIR region, we can see in the figure that we can explain the fitted values with a stars-spike, formed on an underlying generalized NFW profile, with a slope $\gamma \sim 0.8$. As a consequence, we can rule out any DM distribution profile that produces gamma-ray flux above the H.E.S.S. data, with a slope $\gamma \gtrsim 0.8$. When studying the Ridge, Diffuse and Halo regions no profile can be ruled out since the effect of the spike is irrelevant in those regions, meaning that the angular extension for the stars-spike cannot extend in the Diffuse and Ridge region, $\theta_\text{star}\lesssim\theta_\text{Diff} = 0.15^\circ$-$0.45^\circ$ or $R_\text{star} \lesssim R_\text{Diff} = 22$-$65 $ pc, sharing the same spatial constraint as in the adiabatic case.\\

Actually, the star-spike scenario is an appealing one. In fact, the possibility of the existence of a DM density enhancement in a form of an adiabatic spike in the Milky Way remains unclear, mostly due to the uncertainty on its stability and the evolution history of the Galaxy. However, the small dimensions of the possible spike, $R_{\text{sp}} \lesssim 22$-$65$ pc or $\theta \lesssim 0.15^\circ$-$0.45^\circ$, make it very difficult to prove the existence of the DM spike in our galaxy or even in other galaxies so far.

\subsubsection{Effect of the Kerr Black Hole}
\label{J_factors_Kerr}

So far, we have considered in our analysis a non-rotating SMBH. Nonetheless, the recent observation of the Sgr A* shadow by the Event Horizon Telescope shows evidence to conclude that the SMBH at the center of our Galaxy is a rotating Kerr BH \cite{Event_Horizon_Telescope_Collaboration_2022_VI, akiyama2022first}. In fact, the shadow diameter depends on the metric and its properties, e.g. the BH spin. As a result, we investigate whether the SMBH spin could affect the detection of a DM signature. As an example, assuming an Hernquist profile (($\alpha$, $\beta$, $\gamma$) = (1, 4, 1)) an extra maximum factor of 2 appears in the J-factor -depending on the value of the spin- when compared to the non-rotating case. Nonetheless, this extra factor mostly disappears when considering the annihilation factor $\rho_{\text{sat}}$ \cite{Ferrer2017}.

Also, Kerr BH shows asymmetric DM spikes, the asymmetry being dependent on the direction of rotation \cite{Ferrer2017, https://doi.org/10.48550/arxiv.1506.06728}. However, this asymmetry is only important in the inner $\sim 10^{-6}$ pc (or $5R_{\text{S}}$), so it would not be visible by the current gamma-ray telescopes. Another important factor is that the annihilation products are sensitive to the SMBH spin: 1) in general higher spins generate higher-energy annihilation products and 2) the annihilation products from the DM particles close to the BH get redshifted, as this population moves closer and closer to it \cite{https://doi.org/10.48550/arxiv.1506.06728}. Hence, because of the small differences between the rotating and non-rotating cases, for simplicity, we will focus only on the non-rotating case.

\subsection{Instant spike}
\label{J_factors_Instant}

\begin{figure}[t] 
  \centering 
  \includegraphics[width=0.49\textwidth]{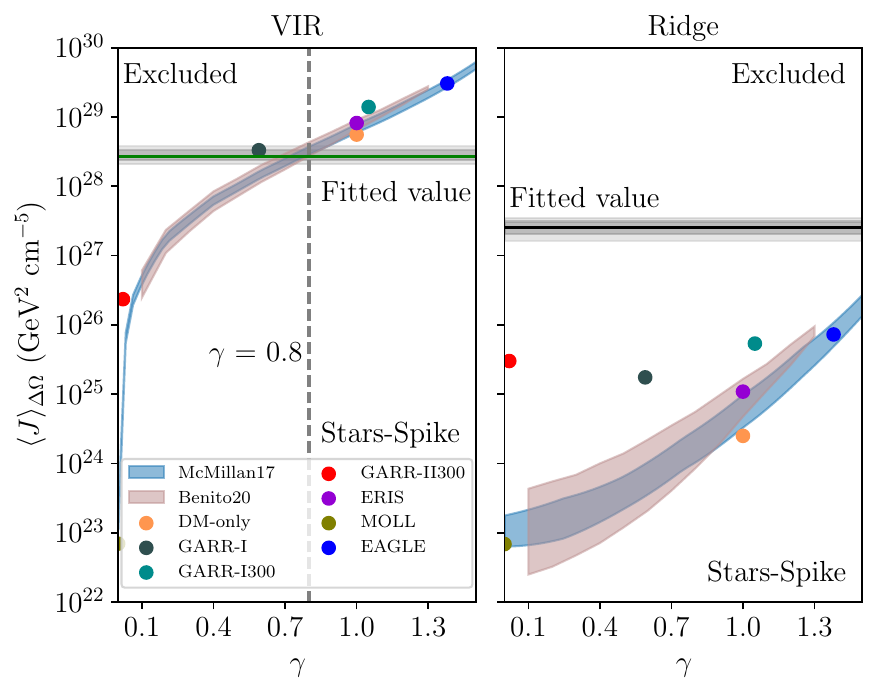} 
  \includegraphics[width=0.49\textwidth]{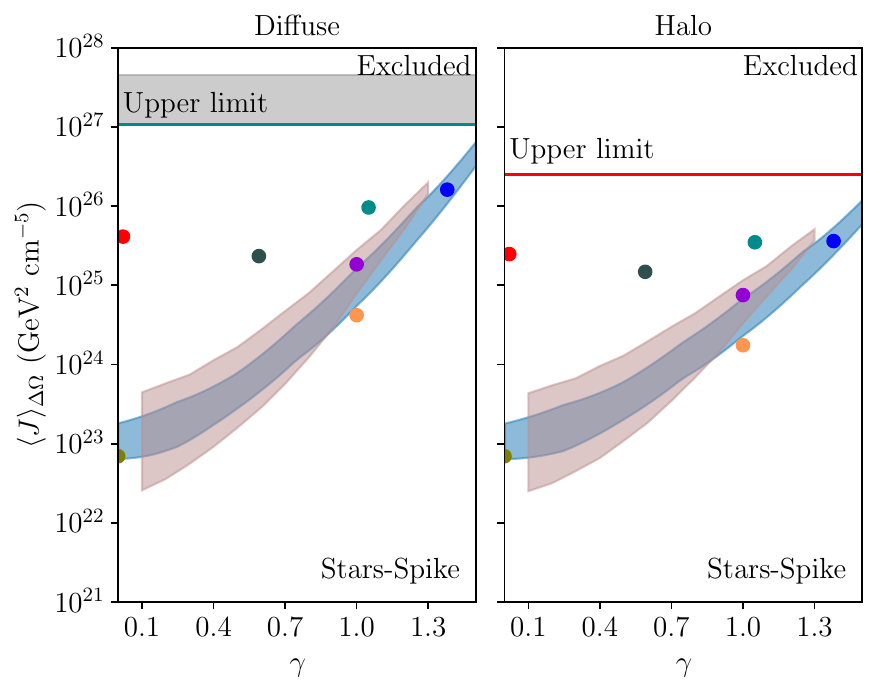} 
  
  \includegraphics[width=0.49\textwidth]{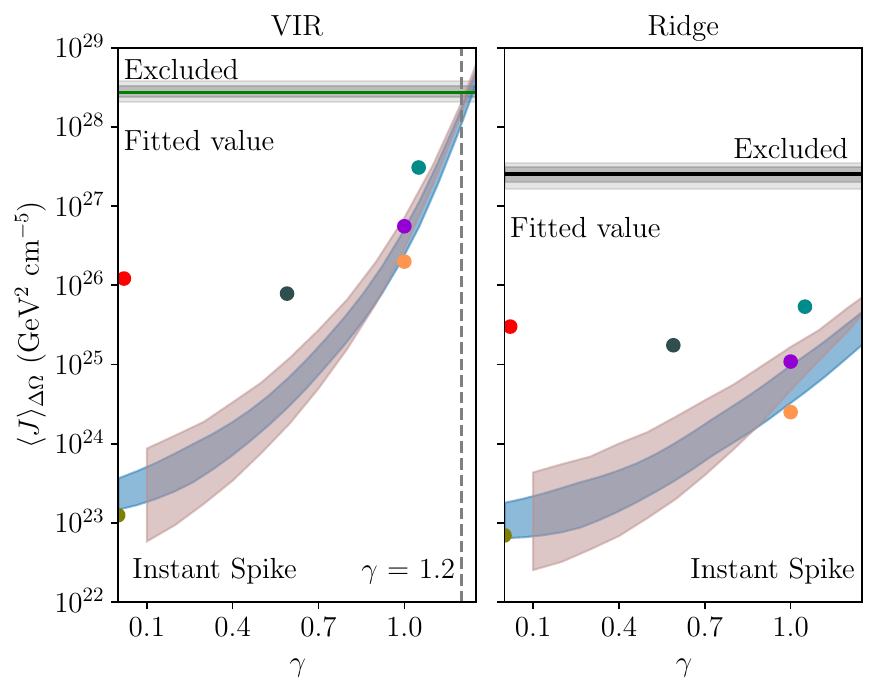} 
  \includegraphics[width=0.49\textwidth]{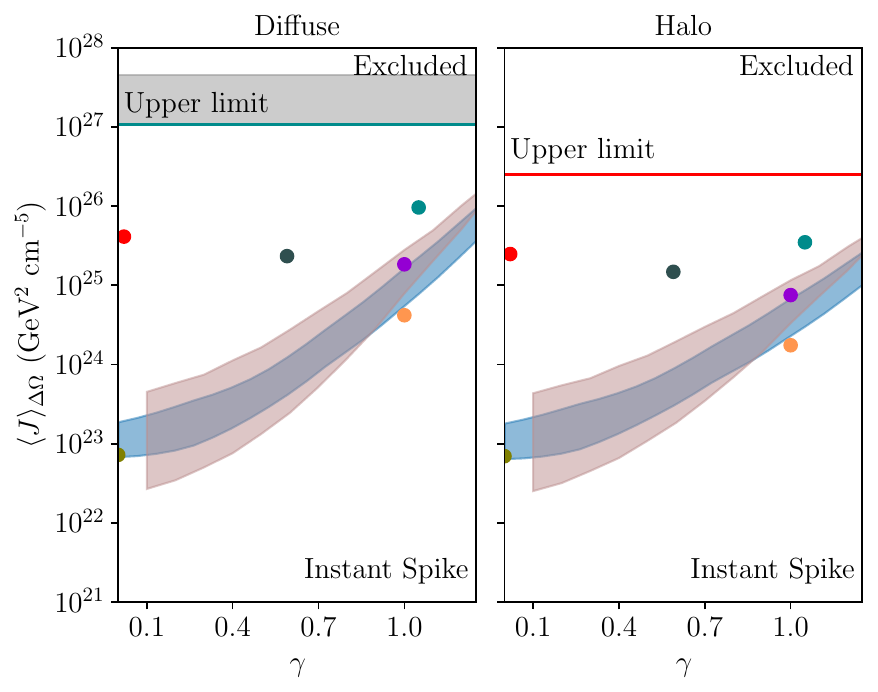} 
   
  \caption{\footnotesize{Same as Figure \ref{fig:JFactors_NFW_Spike}, but now for the approaches stars-spike (first row) and instant spike (second row).} }
  \label{fig:JFactors_stars_instant} 
\end{figure}

Finally, we consider the extreme case that the SMBH is formed instantly as a consequence of a non-adiabatic process \cite{Ullio2001} and the consequences for the DM density profile (dot-dashed line in Figure \ref{fig:DMOnly_all}). In this case, the spike has a less steep profile with an extension of $R_{\text{inst}}$ and slope $\gamma_{\text{inst}}$, where typically have values of $\sim 50$ pc ($\sim 0.15^{\circ}$) and slopes of $\sim 4/3$. The form of the instant spike density profile is given by:

\begin{equation}
  \rho (r) = 
     \begin{cases}
       0 & r < 2 R_{\text{S}}\\
       \rho_{\text{inst}} (r/R_{\text{inst}})^{-\gamma_{\text{inst}}}  &  2 R_{\text{S}} \leq r \leq R_{\text{inst}}\\ 
       \rho_{\text{halo}} (r) & r \geq R_{\text{inst}}\\
     \end{cases}
  \label{eq:DM_profile_instant}
\end{equation}

Where $\gamma_{\text{inst}}$ and $\rho_{\text{inst}}$ are computed numerically following the prescription given in \cite{Ullio2001} for each profile, and $R_{\text{inst}}$ is taken such that $\rho_{\text{inst}} = \rho_{\text{halo}} (R_{\text{inst}})$. In Table \ref{tab:spike_parameter_table} we show the values of the $R_{\text{inst}}$ and $\gamma_{\text{inst}}$ parameters for each profile considered in this work.

We remark that the ``instantaneous'' spike and the ``adiabatic'' spike represent the two relevant opposite limits that have been studied analytically; however, a realistic formation scenario of such DM overdensities may proceed through a non-trivial combination of adiabatic and non-adiabatic phases. A comprehensive description of this process under realistic assumptions is beyond the scope of the present work.

In Figure \ref{fig:JFactors_stars_instant} (lower row) we show the results of the integrated J-factors compared to the fitted values. As we can see, a cuspy profile of $\gamma \sim 1.2$ can explain the fitted J-factors in the VIR region. Note in the instant case only it has been represented $\gamma$ up to 1.25, this is because at higher $\gamma$ the instantaneous spike is very similar to the profile without modification. As we can see, profiles with $\gamma \gtrsim 1.2$ produce gamma-ray flux above the measured H.E.S.S. data in these regions, so we can rule them out. Studying the Ridge, Diffuse and Halo regions from the figure, we see similar results as the rest of the formalisms used (generalized NFW, adiabatic spike or instant spike) with no profile being ruled out. As a conclusion, the same spectral constraint can be extracted as before but now with the instantaneous case: $\theta_{\text{inst}} \lesssim \theta_{\text{Diff}} = 0.15^{\circ}$-$0.45^{\circ}$.

\section{S2 star dynamical constraints}
\label{dynamical_constraints}

In the previous sections, we have presented the constraints on the DM density profile obtained by the spectral study of a collection of gamma-ray observations of the GC by H.E.S.S. and the predictions obtained via both cosmological simulations and galactic dynamics of the outer region of the Galaxy. Nonetheless, we find out that different scenarios could explain the gamma-ray flux observed in the very inner region of the Galaxy, e.g. a more enhanced NFW profile with $\gamma \sim 1.3$ or an adiabatic DM spike formed on a shallower profile. In order to disentangle this dichotomy, we performed another independent analysis with the study of the S2 star orbit \cite{Lacroix_2018, Heissel:2021pcw}.

We can set further constraints on the inner DM density distribution by considering the dynamical constraints obtained by the study of the inner S2 star precession: GRAVITY2020 \cite{GRAVITY_2020}, GRAVITY2021 \cite{GRAVITY_2021} and Do$+2019$ \cite{Do2019}. Indeed, the extended mass within the orbit can modify its predicted precession up to a point where constraints can be extracted. In particular, we update the results obtained in \cite{Lacroix_2018}, by including new observational data (the GRAVITY and Do$+2019$ data).\footnote{During the redaction of this manuscript, \cite{Shen2023} was published with a similar objective as this section. We find that our conclusions are consistent with their results.} 

\begin{figure}[t!] 
  \centering 
  \includegraphics[width=0.49\textwidth]{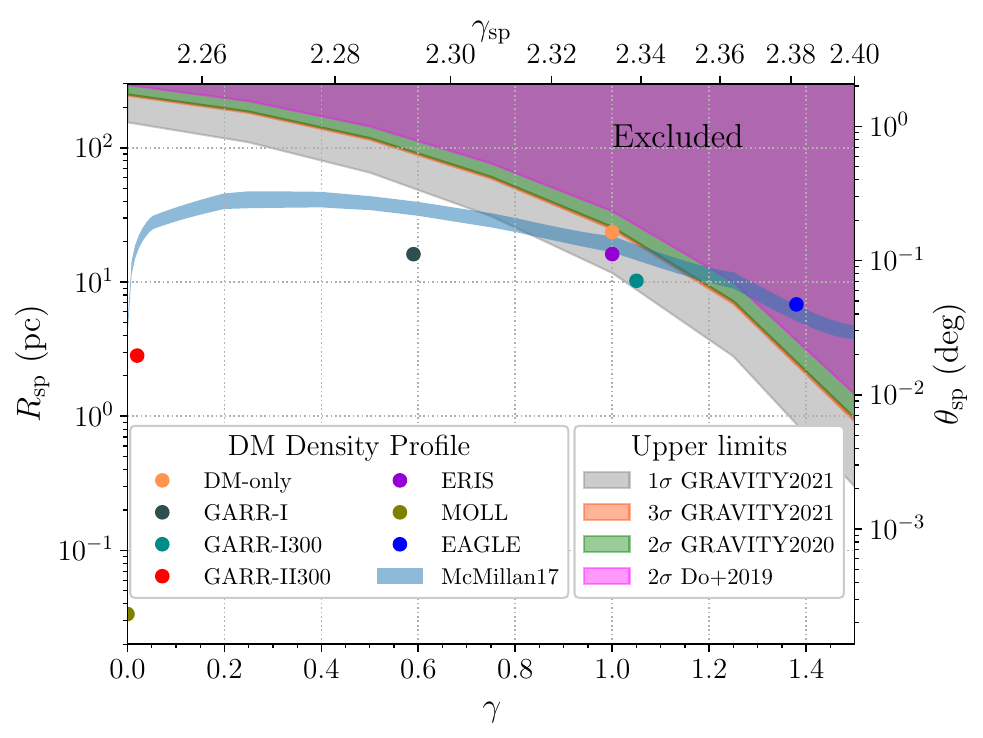} 
  \includegraphics[width=0.49\textwidth]{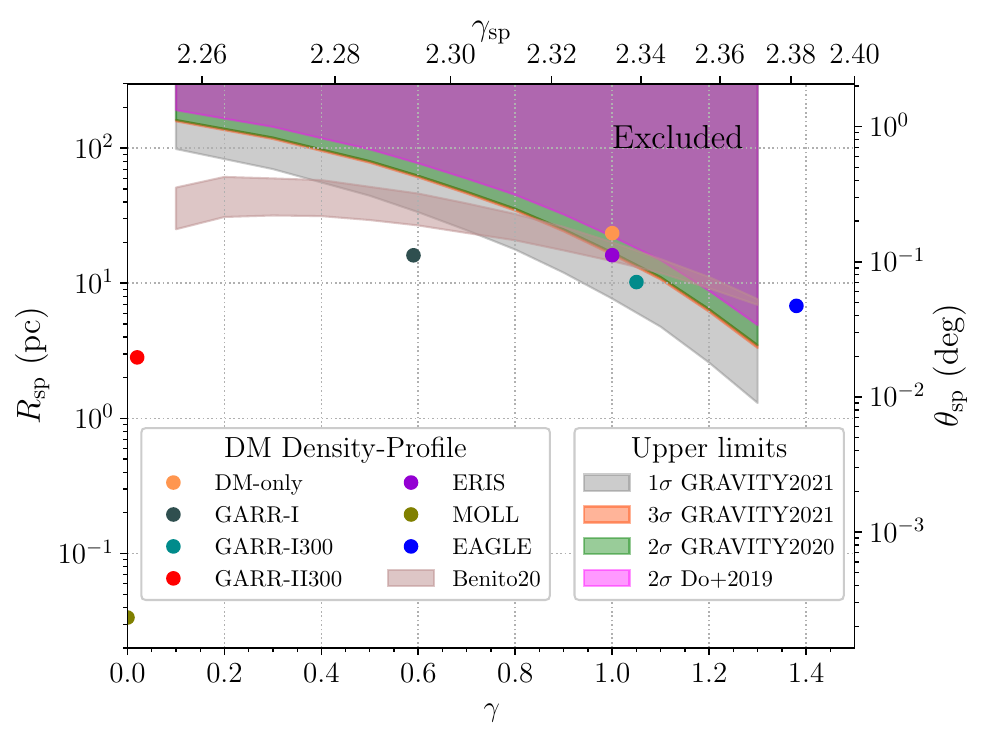} 
  \caption{\footnotesize{Upper limits on the parameter space ($\gamma$, $R_{\text{sp}}$) defined by the slope of the underlying DM profile and the radius of the spike. The exclusion regions are obtained by the study of the S2 orbit and data by: GRAVITY2021 \cite{GRAVITY_2021} (1$\sigma$ in grey region and 3$\sigma$ in orange), GRAVITY2020 \cite{GRAVITY_2020} (2$\sigma$, green region) and by Do$+2019$ \cite{Do2019} (2$\sigma$, purple region). Left panel: upper limits computed with McMillan17 \cite{McMillan2017} as the base model for the adiabatic spike. The blue region represents the integrated profile McMillan17 (with $\alpha = 1$ and $\beta = 3$ fixed) and the dots are the parameters for the simulated DM profiles used in this work. The spike slope is given by $\gamma_{\text{sp}} =  (9 - 2 \gamma)/(4 - \gamma)$. Right panel: same as left panel, but now following the 2$\sigma$ limits of Benito20 \cite{Benito20}, also with $\alpha = 1$ and $\beta = 3$ fixed.}}
\label{fig:Spike_dynamical_constraints} 
\end{figure}

\begin{figure}[t!] 
  \centering 
  \includegraphics[width=0.49\textwidth]{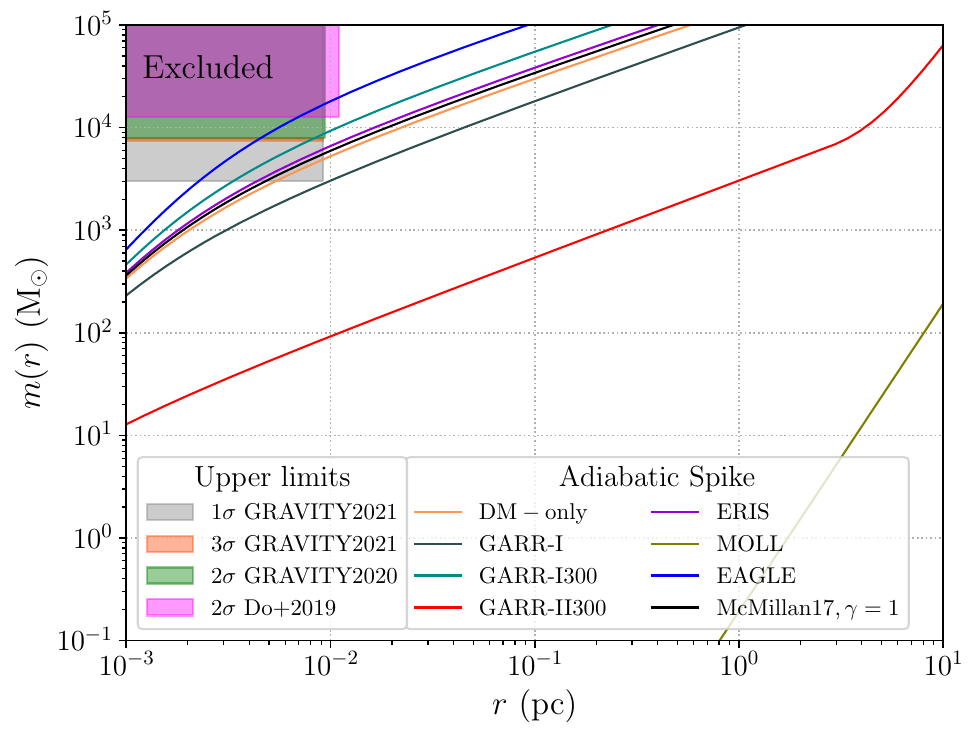} 
  \includegraphics[width=0.49\textwidth]{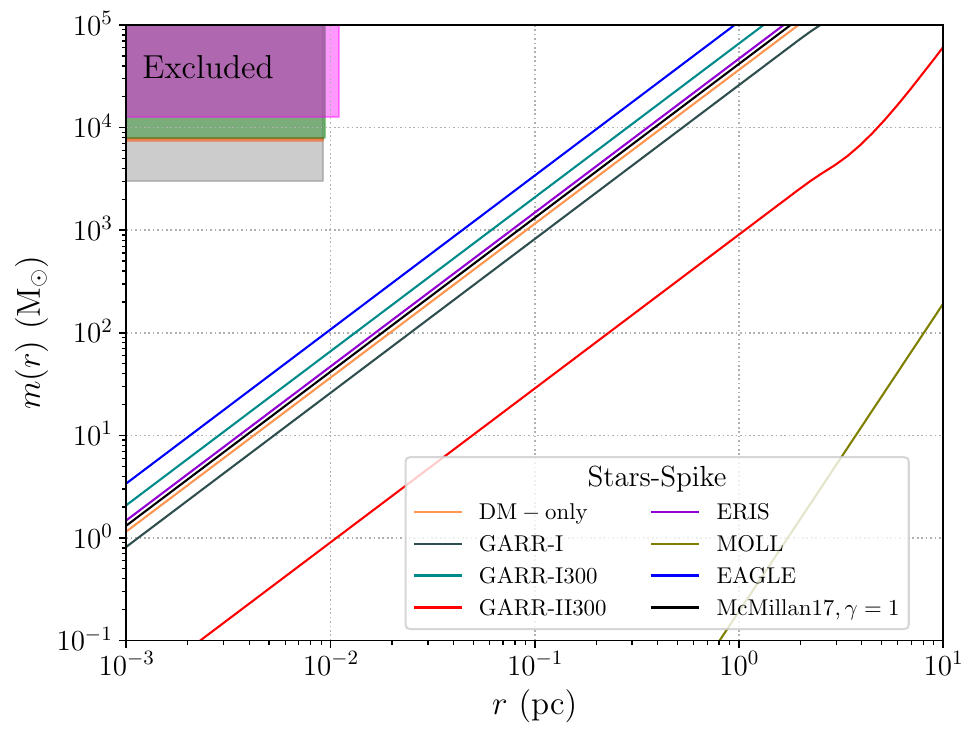} 
  \caption{\footnotesize{Upper limits on the enclosed mass within the S2 orbit, where the excluded regions are given by GRAVITY2021 \cite{GRAVITY_2021} (1$\sigma$ in grey and 3$\sigma$ in orange), GRAVITY2020 \cite{GRAVITY_2020} (2$\sigma$, green region) and by Do$+2019$ \cite{Do2019} (2$\sigma$, purple region). Left panel: integrated mass for all the DM density profiles from cosmological simulations with the adiabatic spike formalism. The excluded profiles are the DM-only, ERIS, GARR-I300 and EAGLE. Right panel: same as left panel, but with the spike-stars approach. No profiles can be excluded this time. The same study has been done for the extrapolated generalized NFW profile and the case of an instant spike, but the results are not shown because no constraints can be put on the models, like in the case of the spike-stars approach. }}
  \label{fig:enclosed_mass} 
\end{figure}

Indeed, we have computed the limits in the parameters ($\gamma_{\text{sp}}$ and $R_{\text{sp}}$) from the mass enclosed within the orbits of the stars, for a DM mass $m_{\text{DM}} = 36$ TeV, in agreement with the best fit of the VIR. In Figure \ref{fig:Spike_dynamical_constraints} we show the exclusion region for such a parameter space. In the left panel, we show the constraints integrating the McMillan17 profile with the BH adiabatic spike prescription given by \cite{PhysRevLett.83.1719, Sadeghian2013} and computing the maximum $R_{\text{sp}}$ allowed by the observations. Note that each upper limit is dependent on the integrated DM profile assumed as a benchmark model, in this case, McMillan17. Within this approach, NFW profiles with $\gamma > 0.75 $-$ 0.85$, or $\gamma_{\text{sp}} > 2.31 $-$ 2.32$, can be ruled out with the current observations and models, leaving the possibility for the existence of adiabatic DM spike formed on shallower profiles, in agreement with the results shown in Figure \ref{fig:JFactors_NFW_Spike} and \ref{fig:JFactors_stars_instant}. A similar procedure is done in the right panel but now taking the 2$\sigma$ region from the values considered in Benito20 as the benchmark model. Note that only values of $\gamma \in$ (0.1, 1.3) are shown because, as it can be seen in the Appendix Figure \ref{fig:params_NFW_vs_gamma}, the 2$\sigma$ values of Benito20 are in this range of $\gamma$. Within this approach, the constraints are slightly more restrictive: $\gamma > 0.40$-$0.75$, or $\gamma_{\text{sp}} > 2.28$-$2.31$, are excluded, keeping a general agreement with the previous case. This leads to the conclusion that if a DM adiabatic spike exists in the GC, necessarily the base model of the DM halo must be flatter than the standard NFW profile ($\gamma \lesssim 0.75$).

In Figure \ref{fig:enclosed_mass} we show the integrated mass for each of the simulation-based DM density models, for the adiabatic spike (left panel) and stars-spike (right panel) approaches, with the limits on the extended mass given by the S2 observations. As we can see, we can rule out the growth of an adiabatic spike on cuspy profiles (DM-only, EAGLE, GARR-I300 and ERIS, all with $\gamma \geq 1$). We do not show the constraints for the rest of the approaches (generalized NFW and instant spike), because no constraints can be set. Indeed, our spectral constraints on the J-factor appear to be more restrictive than the ones coming from the S2 orbit.

\section{Conclusions}
\label{conclusions}

Motivated by previous works, we further investigate the possibility that the gamma-ray spectral cut-off detected by H.E.S.S. at tens of TeV in the Very Inner Region (VIR, $r <15$ pc) could be a DM annihilation signal. Under this hypothesis, we set constraints to the DM density profile in 5 concentric regions around the GC observed by H.E.S.S. (VIR, Ridge, Diffuse, Halo and IGS). In particular, we have considered the hypothesis that the multi-TeV diffuse gamma-ray emission measured by H.E.S.S. in the Diffuse and Ridge regions includes a sub-dominant component associated with DM annihilation (within the WIMP framework), and explored the consequences of this hypothesis on the DM distribution in the very inner region of the Galaxy, with particular focus on the possibility of having a steep profile in the center (sometimes dubbed ``DM spike''). By modeling the expected background in each region with DRAGON and fitting the spectral to the H.E.S.S. data (assuming the thermal relic cross-section), we extract valuable information on the J-factor.\\

The best fit for the DM mass of a thermal WIMP ($\langle\sigma v\rangle \simeq 2.2\times 10 ^{-26} \text{cm}^3 \text{s}^{-1}$) annihilating into ZZ channel in the VIR is $m_{\text{DM}} = 36 ^{+4}_{-6}$ TeV. We assume that the gamma-ray flux generated by the annihilation of TeV DM should also contribute to the gamma-ray flux in the concentric regions. In fact, the radial distribution of the signal-to-noise ratio (between the DM signal and the astrophysical background) strongly depends on both the background modeling and the DM density profile. By fitting the H.E.S.S. gamma-ray data in the different regions, we have determined the amplitude of the DM contribution to the gamma-ray flux, which is related to the DM profile, i.e. the integrated J-factors.\\

Indeed, if this signal is due to a $\sim 36$ TeV thermal WIMP annihilating into in ZZ channel, we can constrain the radial distribution of J-factor from $\sim 10$ pc to $\sim 400 $ pc. The combined analysis of the 5 regions appears to be consistent with the hypothesis, showing an enhancement of the DM density in the inner region with respect to a benchmark NFW profile. In particular, the analysis of gamma-ray data from the VIR is compatible with being originated by a DM signature over the background. We compare our results with different DM profiles obtained by both simulation and dynamical constraints. First of all, to fit the gamma-ray cut-off with the DM component with the assumed thermal relic cross-section, the generalized NFW profile should have a slope $\gamma \sim 1.3$. Interestingly, this result is similar to the one obtained for the Galactic Center Excess observed by Fermi-LAT at the GeV energy scale, which could be explained as a GeV WIMP candidate with an NFW profile with $\gamma =  1.1 $-$ 1.2 $ \cite{Di_Mauro_2021}.\\

\begin{figure}[t!] 
  \centering 
  
  \includegraphics[width=0.49\textwidth]{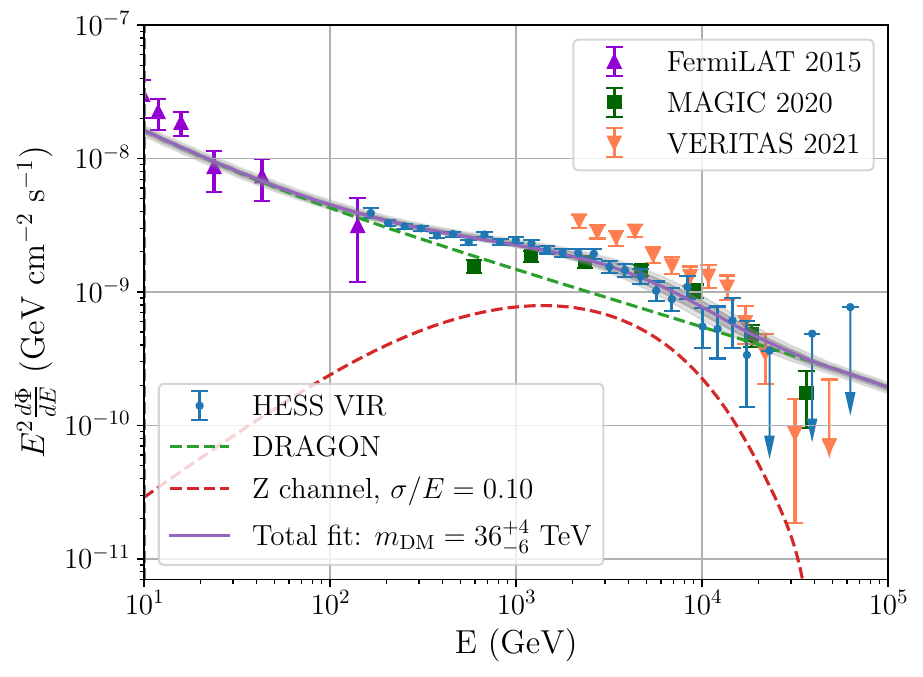}
  \includegraphics[width=0.49\textwidth]{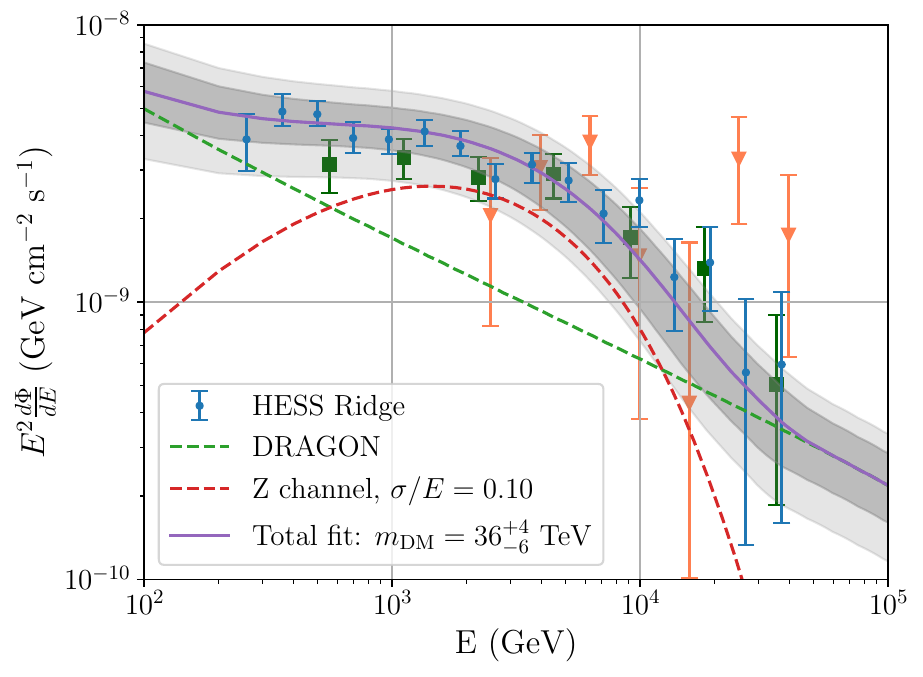}
  \includegraphics[width=0.49\textwidth]{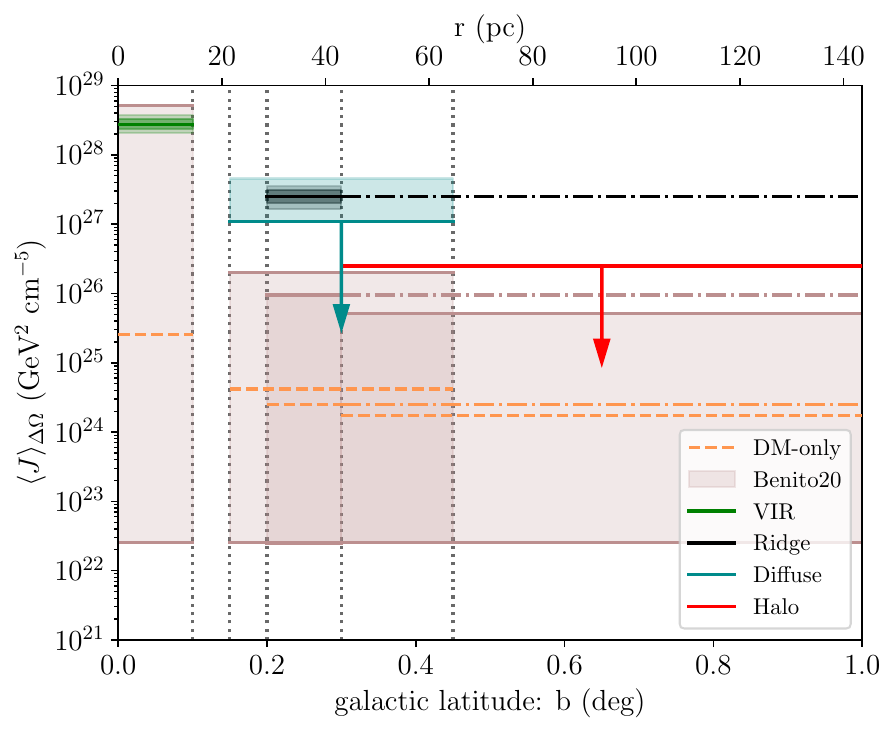} 
  \includegraphics[width=0.49\textwidth]{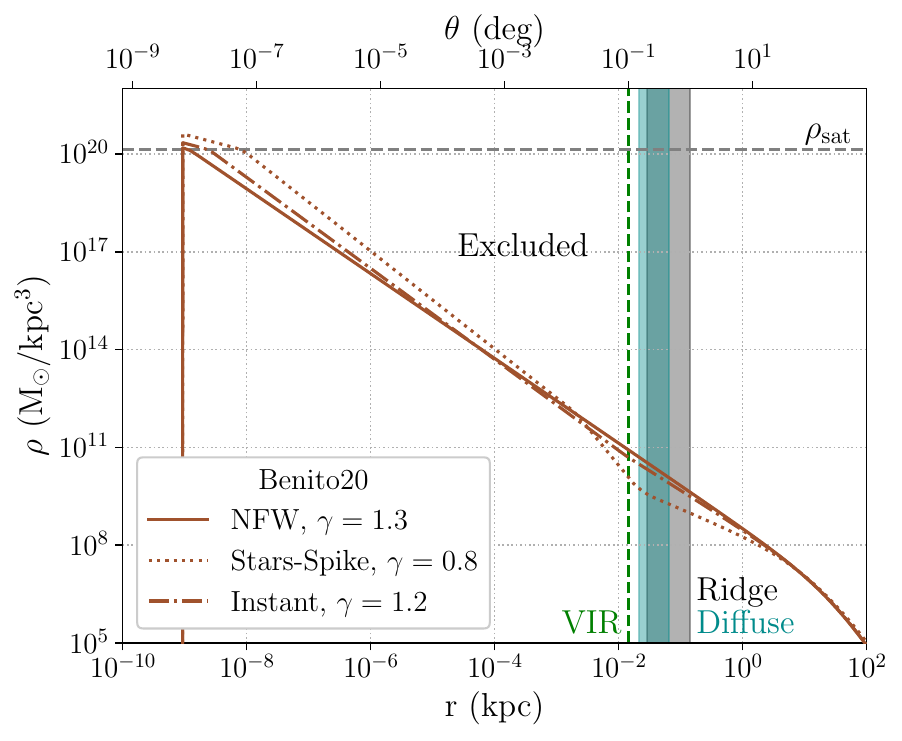} 
  
  \caption{\footnotesize{Upper row: Fermi-LAT, MAGIC and VERITAS data \cite{2021ApJ...913..115A} compared to our fit, showing compatibility for both the VIR (left panel) and Ridge region (right panel). Lower row, left panel: zoom-in of Figure \ref{fig:J_factor_fit_vs_EVANS}. Lower row, right panel: upper limits on the DM density profile (Benito20) coming from the gamma-rays spectral constraints (Figures \ref{fig:JFactors_NFW_Spike} and \ref{fig:JFactors_stars_instant}). We show the profiles with $\gamma = 1.3$ for the generalized NFW, $\gamma = 0.8$ for the Stars-Spike and $\gamma = 1.2$ for the instantaneous spike. The vertical dotted line and colored regions correspond to the spatial extensions of the VIR (in green, $0.1^\circ$ or 15 pc), Diffuse (in blue, $0.15^\circ$-$0.45^\circ$ or 22-65 pc) and Ridge (in gray, $0.2^\circ$-$1.0^\circ$ or 29-145 pc).}}
  \label{fig:best_profiles} 
\end{figure}

We can exclude generalized NFW profiles with $\gamma \gtrsim 1.3$, which would produce more gamma-ray flux than that observed and, also, is in agreement with estimates of the DM density profile in the Galactic bulge \cite{2017PDU....15...90I}. Further, we consider the possibility of having different underlying DM profiles, whose inner slopes could also be modified in the inner region by the presence of a DM spike associated with the SMBH Sgr A*. Depending on the slope and size of the spike (which also depends on the characteristics of the original underlying DM profile and spike formation history), we can exclude several scenarios that involve a strong overdensity in the inner GC region. In particular, we find out that if the SMBH Sgr A* grew adiabatically, a DM spike could be formed only in cored profiles ($\gamma \sim 0$) to be compatible with our spectral analysis, implicitly pointing to a different origin for the gamma-ray data in the VIR. This scenario seems to disfavor the existence of a pure adiabatic DM spike in the GC. When including the interactions of the stars with the adiabatic spike, we can exclude any underlying profile with $\gamma \geq 0.8$. For the extreme case of the instant spike, the excluded slope is $\gamma \geq 1.2$. Finally, in more external regions (Ridge, Diffuse, Halo and IGS) the gamma-ray data and upper limits are consistent with the absence of a detectable signal and, indeed, with any outer DM distribution among those considered here. See Table \ref{tab:final_results} to see a summary of the results of this paper.\\

\begin{table}[t!]
    \begin{center}
        \begin{tabular}{|c|c|c|c|c|}
\hline 
\hline
Profile  & VIR  & Ridge & Diffuse & S2-star \\ 
\hline 
\hline
Gen. NFW     & Excluded $\gamma \gtrsim 1.28\pm 0.02$     & Allowed     & Allowed       & Allowed \\ \hline
Adiabatic           & Excluded                          & Allowed      & Allowed       & Excluded  $\gamma \gtrsim 0.6 \pm 0.2$ \\ \hline
Star-spike          & Excluded $\gamma \gtrsim 0.76^{+0.04}_{-0.07}$     & Allowed      & Allowed       & Allowed \\ \hline
Instant             & Excluded $\gamma \gtrsim 1.22\pm 0.02$     & Allowed     & Allowed       & Allowed \\ 
\hline
 \end{tabular}
        \caption{\footnotesize{Final remarks of this work. The DM distribution in the Galaxy could be described by a Generalized NFW profile with $\gamma \lesssim 1.3$ till the innermost part of the Galaxy. Adiabatic DM spikes are mainly excluded by both spectral analyses of the inner 10 pc of the Galaxy and dynamical constraints by S2-star, at least of an interaction with baryonic matter which smooths the steepness of the profiles. The latter case would allow star-spike profiles with $\gamma \gtrsim 0.8$. Instant DM spike would be compatible with the data for $\gamma \lesssim 1.2$. The uncertainty corresponds to a 2$\sigma$ confidence level.}}
        \label{tab:final_results}
    \end{center}
\end{table}

By the analysis of the Diffuse region, which does not include the VIR and where the DM contribution to the gamma-ray signal is already no longer significant, we can conclude that the angular extension of the DM density enhancement that is required to explain the gamma-ray data is $\theta_{\text{sp}} \lesssim \theta_{\text{Diff}} = 0.15^{\circ} $-$ 0.45^{\circ}$, i.e. $R_{\text{sp}} \lesssim R_{\text{Diff}} = 22 $-$ 65 \text{ pc}$.  Interestingly, this result is in agreement with the astrophysical interpretation of the Galactic center excess detected by the Fermi-LAT satellite. On the one hand, our renormalization of DRAGON model for the VIR region is in agreement with the Fermi-LAT data with $\text{E} \gtrsim 10$ GeV (Figure \ref{fig:best_profiles}, upper left panel), where the Fermi-LAT angular resolution is comparable with the VIR region. On the other hand, the spatial resolution of Fermi-LAT deteriorates at $\text{E}\lesssim 10$ GeV, making feasible the spatial inclusion of other components in the spectral emission.\\

In Figure \ref{fig:best_profiles} (upper right panel) we compare our fit of the Ridge region with Fermi-LAT, MAGIC and VERITAS data \cite{2021ApJ...913..115A}. Although the gamma-ray spectra is well fitted by our model, the best fit for the J-factor favors the hypothesis of an astrophysical gamma-ray emission for this region. This is consistent with the fact that the Ridge region is a non-spherical one, including up $1^\circ$ deg of the Galactic plane. This fact is appreciable in the lower left panel of Figure \ref{fig:best_profiles}: the J-factor of the Ridge region - if due to a DM component - is expected to be lower than the J-factor in the Diffuse. Indeed, we can conclude that the gamma-ray emission at a longitude greater than $0.2^\circ$ from the Galactic center is dominated by the astrophysical background.

Finally, we verify the compatibility of our results with the upper limits on the total DM mass at the sub-parsec scale obtained by the independent analysis of the S2 orbit\cite{GRAVITY_2020, Do2019, GRAVITY_2021}. Depending on the assumed DM density distribution profile we are able to rule out profiles with $\gamma > 0.75 $-$ 0.85$ (model: McMillan17, see text for more details) and $\gamma > 0.40$-$0.75$ (model: Benito20, see text for further details), in the case that a DM spike is formed adiabatically. This leads to the conclusion that, if a DM adiabatic spike exists in the GC, necessarily the underlying DM halo would be more shallow than the benchmark NFW profile. This result is independent and complementary to the gamma-ray analysis, although it is less constraining. On the one hand, beyond the adiabatic spike, a less spiky profile cannot be excluded from the study of S2 orbit. On the other hand, the adiabatic spike is already excluded from the cuspy-like profile by the spectral analysis of this work.

To conclude, our analysis seems to discard the possibility of having an adiabatic DM spike in the GC. Also, it reinforces the possibility of a DM profile with slope $\gamma \sim 1.3$. The latter possibility should be further checked with an in-depth study of a region between $0.1^{\circ}-1^{\circ}$ deg, where the DM density is expected to increase gradually (Figure \ref{fig:best_profiles}, lower right panel).  Our work represents a proof-of-concept paper in order to guide the study of the GC with both current and next-generation of gamma-ray telescopes. In particular, the Cherenkov Telescope Array (CTA) \cite{Acharyya_2021} will be able to scan the GC region with a flux sensitivity of $\sim 1$ order of magnitude better than H.E.S.S. and angular resolution of $\sim 0.05^{\circ}$-$0.03^{\circ}$, instead of the $\sim 0.1^{\circ}$ of H.E.S.S.

\acknowledgments
The work of MASC, JZP and VG were supported by the grants PID2021-125331NB-I00 and CEX2020-001007-S, funded by MCIN/AEI/10.13039/501100011033, by ``ERDF A way of making Europe'', and the MULTIDARK Project RED2022-134411-T. JZP's contribution to this work has been supported by \textit{FPI Severo Ochoa} PRE2021-099137 grant and "\textit{Ayudas para el fomento de la investigación en Estudios de Máster-UAM 2021}". VG's contribution to this work has been supported by \textit{Juan de la Cierva-Incorporaci{\'o}n} IJC2019-040315-I grants. VG and JZP are also supported by the grant PID2022-139841NB-I00 and thank Rafael Alvares Batista and all the DAMASCO group for useful discussions. VG also thanks Fabio Iocco for useful discussions. DG acknowledges support from the project “Theoretical Astroparticle Physics (TAsP)” funded by the INFN.

\appendix
\section{Parameters of the DM density profile}
\label{Appendix_params}

\begin{figure}[t!] 
  \centering 
  \includegraphics[width=0.49\textwidth]{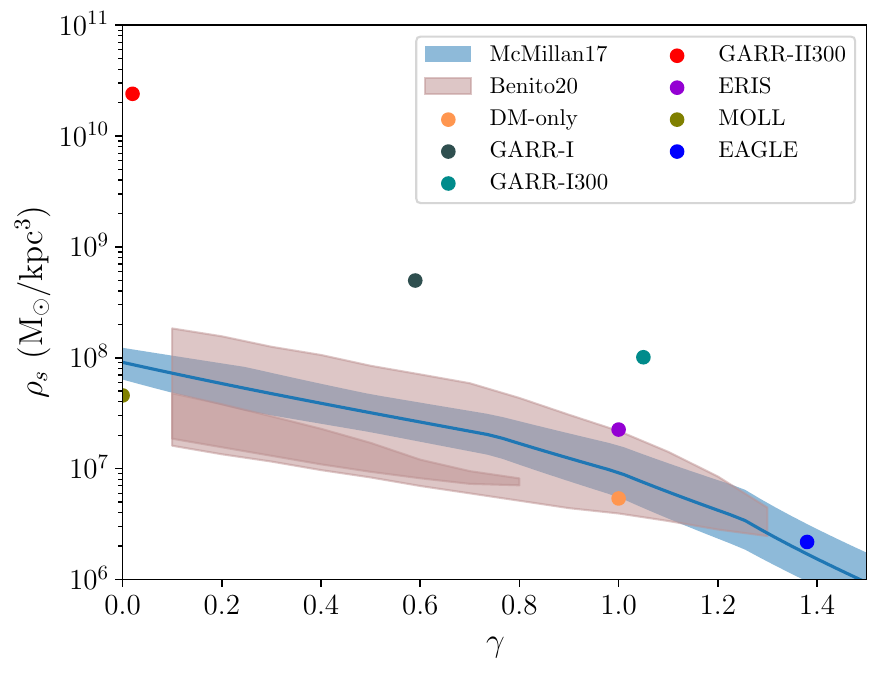} 
   \includegraphics[width=0.49\textwidth]{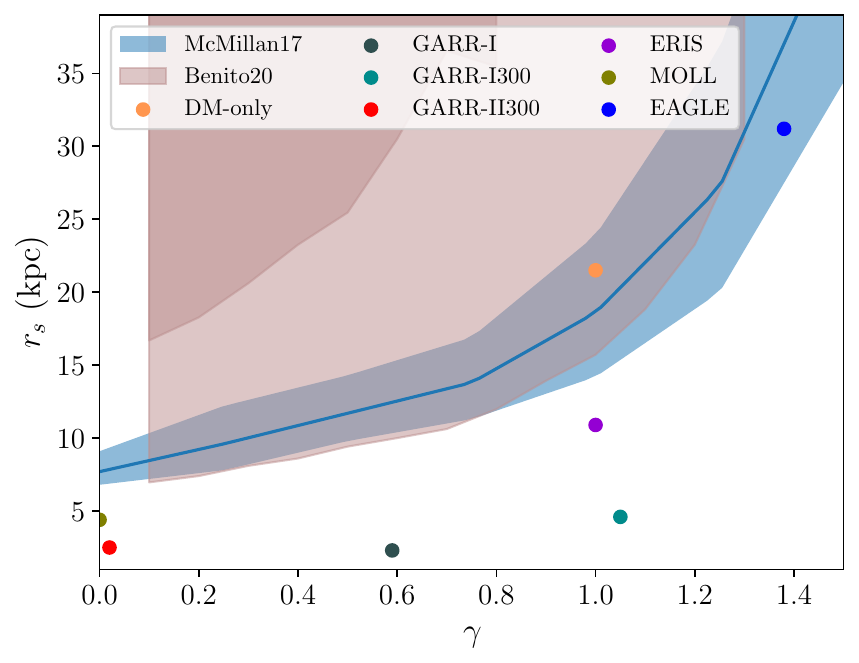} 
    \caption{\footnotesize{Parameters of the different DM density profiles used in this work. The dots correspond to the values given by numerical simulations (Table \ref{tab:density_parameter_table}) and the blue and brown regions correspond to the McMillan17 \cite{McMillan2017} and Benito20 \cite{Benito20} models (1$\sigma$ and 2$\sigma$ regions), given by observations.}}
  \label{fig:params_NFW_vs_gamma} 
\end{figure}

In Figure \ref{fig:params_NFW_vs_gamma} of this Appendix we give the parameters of the generalized NFW profile defined in Equation \ref{eq:NFW_general} of the main text:

\begin{equation}
  \rho_{\text{halo}}(r) = \frac{ \rho _s}{(\frac{r}{r_{s}})^{\gamma} (1+(\frac{r}{r_{s}})^{\alpha})^{\frac{\beta - \gamma}{\alpha}} } 
  \tag{\ref{eq:NFW_general}}
\end{equation}

In the figure, we show the scale parameters of the DM density distribution profile of a Milky-Way-like galaxy, obtained by N-body (DM-only) and hydrodynamical (GARR-I, GARR-I300, GARR-II300, ERIS, MOLL and EAGLE) simulations as defined in \cite{Gammaldi_2016}. Furthermore, we consider the observational models McMillan17 \cite{McMillan2017} and Benito20 \cite{Benito20}, where the $\alpha$ and $\beta$ parameters of both observational models are $\alpha = 1$, $\beta = 3$. Throughout the paper, we use the 2$\sigma$ region of the Benito20 frequentist approach, being the most conservative. 

\begin{figure}[t!] 
  \centering 
  \includegraphics[width=9cm]{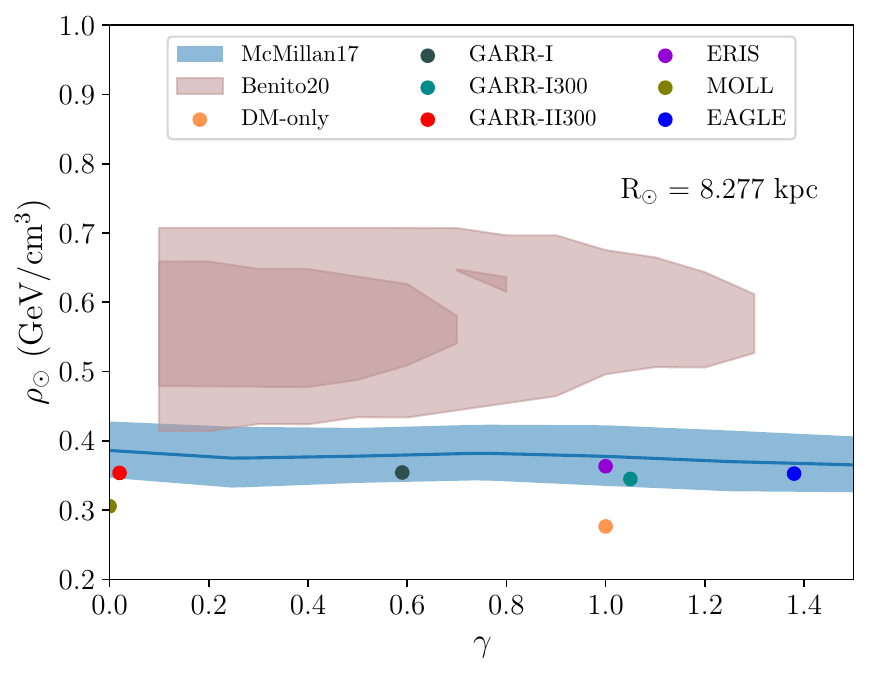} 
  \caption{\footnotesize{Same as Figure \ref{fig:params_NFW_vs_gamma}, but for the local DM density, with $R_{\odot} = 8.277$ kpc (GRAVITY2021 \cite{GRAVITY_2021}).}}
  \label{fig:rho_sun_vs_gamma} 
\end{figure}

All the different DM distribution models are renormalized and compatible with each other. In Figure \ref{fig:rho_sun_vs_gamma} we show the value of the local DM density with $R_\odot = 8.277$ kpc (GRAVITY2021 \cite{GRAVITY_2021}), the position used to compute the J-factors along this work. As it can be seen in the figure, Benito20 gives a higher local DM density than McMillan17 and the rest of the numerical models.

\bibliographystyle{JHEP}
\bibliography{bibliography}

\end{document}